\documentclass{aa}  

\def\4u{4U1820-303}

\def\cstat{{\sc C-stat}}

\def\nh{{$N_{\rm H}$}} 
\def\be{\begin{equation}} 
\def\ee{\end{equation}}

\def\ergscm2{erg s$^{-1}$ cm$^{-2}$}
\def\xspec{\texttt{XSPEC}}
\def\specgauss{Spectrum$_{20000}$counts}
\def\specpoiss{Spectrum$_{2000}$counts}
\def\jaxspec{\texttt{jaxspec}}
\def\sixsa{\texttt{SIXSA}}

\usepackage{graphicx}
\usepackage{txfonts}
\usepackage{epstopdf}
\begin{document}

   \title{Simulation-Based Inference with Neural Posterior Estimation applied to X-ray spectral fitting}

   \subtitle{Demonstration of working principles down to the Poisson regime}
    \titlerunning{Neural networks for X-ray spectral fitting}
   \author{Didier Barret
          \inst{1}
          \and
          Simon Dupourqué\inst{1}}

   \institute{Institut de Recherche en Astrophysique et Planétologie, 9 avenue du Colonel Roche, Toulouse, 31028, France\\
              \email{dbarret@irap.omp.eu}
              }

   \date{Received January 6th, 2024; accepted February, 19th, 2024}

 
  \abstract
   {Neural networks are being extensively used for modeling data, especially in the case where no likelihood can be formulated.}
   {Although in the case of X-ray spectral fitting, the likelihood is known, we aim to investigate the neural networks ability to recover the model parameters but also their associated uncertainties, and compare its performance with standard X-ray spectral fitting, whether following a frequentist or Bayesian approach.}
   {We apply Simulation-Based Inference with Neural Posterior Estimation (SBI-NPE) to X-ray spectra. We train a network with simulated spectra generated from a multi-parameter source emission model folded through an instrument response, so that it learns the mapping between the simulated spectra and their parameters and returns the posterior distribution. The model parameters are sampled from a predefined prior distribution. To maximize the efficiency of the training of the neural network, yet limiting the size of the training sample to speed up the inference, we introduce a way to reduce the range of the priors, either through a classifier or a coarse and quick inference of one or multiple observations. For the sake of demonstrating  working principles, we apply the technique to data generated from and recorded by the NICER X-ray instrument: a medium resolution X-ray spectrometer, covering the 0.2-12 keV band. We consider here simple X-ray emission models with up to 5 parameters.}
   {SBI-NPE is demonstrated to work equally well as standard X-ray spectral fitting, both in the Gaussian and Poisson regimes, both on simulated and real data, yielding fully consistent results in terms of best fit parameters and posterior distributions. The inference time is comparable to or smaller than the one needed for Bayesian inference, when involving the computation of large Markov Chain Monte Carlo chains to derive the posterior distributions. On the other hand, once properly trained, an amortized SBI-NPE network generates the posterior distributions in no time (less than 1 second per spectrum on a 6-core laptop). We show that SBI-NPE is less sensitive to local minima trapping than standard fit statistic minimization techniques. With a simple model, we find that the neural network can be trained equally well on dimension-reduced spectra, via a Principal Component Decomposition, leading to a faster inference time with no significant degradation of the posteriors.}
   {We have shown that simulation-based inference with neural posterior estimation adds up as a complementary tool for X-ray spectral analysis. The technique is robust and produces well calibrated posterior distributions. It holds great potential for its integration in pipelines developed for processing large data sets. The code developed to demonstrate the first working principles of the technique introduced here is released through a Python package called \sixsa\ (Simulation-based Inference for X-ray Spectral Analysis), available from GitHub.}

   \maketitle
%

\section{Introduction}
X-ray spectral fitting relies generally on frequentist and Bayesian approaches; see \cite{Buchner2023arXiv230905705B} for a recent and comprehensive review on the statistical aspects of X-ray spectral analysis. Fitting of X-ray spectra with neural networks has been introduced by \cite{Ichinohe2018MNRAS.475.4739I} for the analysis of high spectral resolution galaxy cluster spectra and recently by \cite{Parker2022MNRAS.514.4061P} for the analysis of lower resolution Athena Wide Field Imager spectra of Active Galactic Nuclei. \cite{Parker2022MNRAS.514.4061P} showed that neural networks delivered comparable accuracy to spectral fitting, while limiting the risk of outliers caused by the fit getting stuck into a local false minimum (the nightmare of anyone involved in X-ray spectral fitting), yet providing an improvement of around three orders of magnitude in speed, once the network had been properly trained. On the other hand, no error estimates on the spectral parameter were provided in the methods explored by \cite{Parker2022MNRAS.514.4061P}. 

\begin{figure*}[!t]
    \begin{center}
    \includegraphics[width=0.9\textwidth]{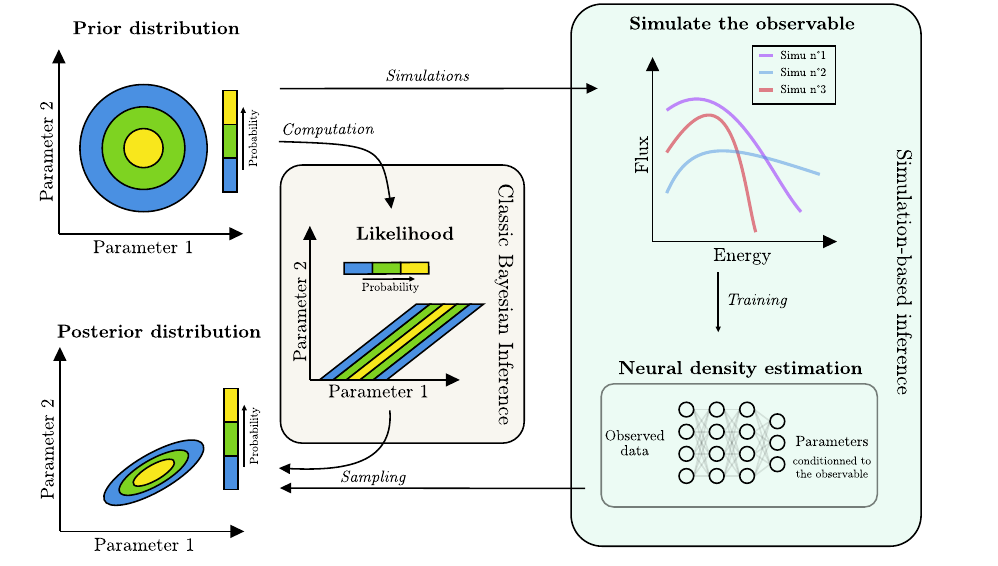}
    \end{center}
    \caption{The Simulation-Based Inference approach emulates the traditional Bayesian inference approach. When assessing the parameters of a model, one first defines prior distributions, then defines the likelihood of a given observation, often using a forward-modeling approach. This likelihood is further sampled to obtain the posterior distribution of the parameters. The simulation-based approach does not require explicit computation of the likelihood, and instead will learn an approximation of the desired distribution (i.e. the likelihood or directly the posterior distribution) by training a neural network with a sample of simulated observations.}
    \label{fig:db_sd_f1}
\end{figure*}

However, in a Bayesian framework, accessing the posterior distribution is possible through the simulation-based inference with amortized neural posterior estimation methodology;  hereafter SBI-NPE \citep{Papamakarios2016arXiv160506376P,Lueckmann2017arXiv171101861L,Greenberg2019arXiv190507488G,Deistler2022arXiv221004815D}, see also \cite{Cranmer2020PNAS..11730055C} for a review of simulation-based inference. In this approach, we sample parameters from a prior distribution and generate synthetic spectra from these parameters. Those spectra are then fed to a neural network that learns the association between simulated spectra and the model parameters. The trained network is then applied to data, to derive the parameter space consistent with the data and the prior, being the posterior distribution. In contrast to conventional Bayesian inference, SBI is also applicable when one can run model simulations, but no formula or algorithm exists for evaluating the probability of data given the parameters, i.e. the likelihood.

SBI-NPE has demonstrated its power in many fields, including astrophysics, e.g. to cite a few, for reconstructing galaxy spectra and inferring their physical parameters \citep{Khullar2022MLS&T...3dLT04K}, for inferring variability parameters from dead-time-affected light curves \citep{Huppenkothen_2022MNRAS.511.5689H}, for exoplanet atmospheric retrieval \citep{Vasist2023A&A...672A.147V}, for deciphering the ring down phase signal of the black hole merger GW150914 \citep{Crisostomi2023PhRvD.108d4029C} and very recently for isolated pulsar population synthesis \citep{Graber2023arXiv231214848G}.

In this paper, we demonstrate for the first time the power of SBI-NPE for X-ray spectral fitting, to show that it delivers performances fully consistent with the \xspec\ \citep{Arnaud1996ASPC..101...17A} and the Bayesian X-ray Analysis (BXA) spectral fitting packages \citep{Buchner2014A&A...564A.125B}; two of the most commonly used tools for X-ray fitting. The paper is organized as follows. In Sect.~\ref{sbi_formalism}, we give some more insights on the SBI-NPE method. In Sect.~\ref{restricted_prior}, we present the methodology to produce the simulated data, introducing a method to reduce the prior range. In Sect.~\ref{Single round inference}, we show examples of single round inference in the Gaussian and Poisson regimes for simulated mock data. In Sect.~\ref{Multiple round inference}, we present a case based on multiple round inference. In Sect.~\ref{Sensitivity to local minima}, we demonstrate the robustness of the technique against local minima trapping. In Sect.~\ref{Dimension reduction with the Principal Component Analysis}, using a simple model, we apply the Principal Component Analysis to reduce the data fed to the network. In Sect.~\ref{Application to real data}, we show the performance of SBI-NPE on real data, as recorded by the NICER X-ray instrument \citep{Gendreau2012SPIE.8443E..13G}. In Sect.~\ref{Discussion}, we discuss the main results of the paper, listing some avenues for further investigations. This precedes a short conclusion.

\section{SBI with amortized neural posteriors}
\label{sbi_formalism}
\subsection{Formalism}

The SBI approach, illustrated in Fig.~\ref{fig:db_sd_f1}, aims at computing the probability distribution of interest, in this case the posterior distribution $p(\vec{\theta}|\vec{x})$, by learning an approximation of the probability density function from a joint sample of parameters $\left\{\vec{\theta}_i\right\}$ and the associated simulated observable $\left\{\vec{x}_i\right\}$, using neural density estimators such as normalizing flow. A normalizing flow $T$ is a diffeomorphism between two random variables, say $\vec{X}$ and $\vec{U}$, which links their following density functions as follows: 

$$ \vec{X} = T(\vec{U}); \quad p_{\vec{X}}(\Vec{x}) = p_{\vec{U}}(\Vec{u}) \left| \text{det} ~ J_{T}(\Vec{u})\right|^{-1} $$

where $J_T$ is the Jacobian matrix of the normalizing flow. The main idea when using normalizing flows is to define a transformation between a simple distribution (i.e. normal distribution) and the probability distribution that should be modelled, which eases the manipulation of such functions. To achieve this, an option is to compose several transformations $T_i$ to form the overall normalizing flow $T$, each parameterized using Masked Auto-encoders for Density Estimation \citep[MADE,][]{germainMADEMaskedAutoencoder2015}, which are based on deep neural networks. MADEs satisfy the auto-regressive properties necessary to define a normalizing flow and can be trained to adjust to the desired probability density. Stacking several MADEs will form what is defined as a Masked Auto-regressive Flow \citep[MAF,][]{papamakariosMaskedAutoregressiveFlow2017}. We refer interested readers to the following reviews by \cite{papamakariosNormalizingFlowsProbabilistic2021, kobyzevNormalizingFlowsIntroduction2021}. \cite{greenbergAutomaticPosteriorTransformation2019} developed a methodology which enabled the use of MAFs to directly learn the posterior distribution of a Bayesian inference problem, using a finite set of parameters and associated observable $\left\{\vec{\theta}_i, \vec{x}_i\right\}$. Using this approach, one can compute an approximation for the posterior distribution $q(\vec{\theta}, \vec{x}) \simeq p(\vec{\theta}| \vec{x})$, which can be used to obtain samples of spectral model parameters from the posterior distribution conditioned on an observed X-ray spectrum.


The python scripts from which the results presented here use the \texttt{sbi}\footnote{https://sbi-dev.github.io/sbi/} package \citep{Tejero2020JOSS....5.2505T}. \texttt{sbi} is a PyTorch-based package that implements SBI algorithms based on neural networks.  It eases inference on black-box simulators by providing a unified interface to state-of-the-art algorithms together with very detailed documentation and tutorials. It is straightforward to use, involving the call of just a few Python functions. 

Amortized inference enables the evaluation of the posterior for different observations without having to re-run inference. On the other hand, multi-round inference focuses on a particular observation. At each round, samples from the obtained posterior distribution computed at the observation are used to generate a new training set for the network, yielding in a better approximation of the true posterior at the observation. Although fewer simulations are needed, the major drawback is that the inference is no longer be amortized, being specific to an observation. We will try both approaches in the sections below.

\subsection{Bench-marking against the known likelihood}
SBI implements machine learning techniques in situations where the likelihood is undefined, hampering the use of conventional statistical approaches. In our case, the likelihood is known. Here we recall the basic equations. Taking the notation of \xspec\ \citep{Arnaud1996ASPC..101...17A}, the likelihood of Poisson data (assuming no background) is known as :
\begin{equation}
	L=\prod_{i=1}^{n}(tm_i)^{S_i}e^{-tm_i}/S_i!
\end{equation}
where $S_i$ are the observed counts in the bin $i$ as recorded by the instrument, $t$ the exposure time over which the data were accumulated, and $m_i$ the predicted count rates based on the current model and the response of the instrument folding in, its instrument efficiency, its spectral resolution, its spectral coverage… see \citep{Buchner2023arXiv230905705B} for details on the folding process. The associated negative log-likelihood, given in \cite{Cash1979ApJ...228..939C} and often referred to as the Cash-statistic, is:
\begin{equation}
C = 2\sum_{i=1}^N (tm_i) - S_i \ln{(tm_i)} + \ln{(S_i!)}
\end{equation}
The final term which depends exclusively on the data (and hence does not influence the best-fit parameters) is replaced by its Stirling's approximation to give :
\begin{equation}
C = 2\sum_{i=1}^N (tm_i) - S_i + S_i (\ln{(S_i)} - \ln{(tm_i)})
\end{equation}
This is what is used for the statistic \cstat\ option in \xspec\: the best fit model is the one that leads to the lowest \cstat\footnote{In the Gaussian, high count regime, the $\chi^2$ statistic is often used in the fitting, but here we used \cstat\ as it applies both in the Gaussian and Poisson regimes without any bias at low counts, e.g. \cite{Buchner2023arXiv230905705B}.}. The default \xspec\ minimization method uses the modified Levenberg-Marquardt algorithm based on the \texttt{CURFIT} routine from \cite{bevington2003data}. In the following sections, we use \xspec\ with and without Bayesian inference, and compute Markov Chain Monte Carlo (MCMC) chains to get the parameter probability distribution and to compute errors on the best fit parameters, as to enable a direct comparison with the posterior distributions derived from SBI-NPE. By default, we use the Goodman-Weare algorithm \citep{2010CAMCS...5...65G}, with 8 walkers, a burn-in phase of 5000 and a length of 50000. The analysis was performed with the \texttt{pyxspec} wrapper of \xspec\ v.12.13.1 \citep{Arnaud1996ASPC..101...17A}.

In addition to \xspec\, we have used the Bayesian X-ray Analysis (BXA) software package \citep{Buchner2014A&A...564A.125B} for the validation of our results. Among many useful features, BXA connects \xspec\ to the nested sampling algorithm as implemented in \texttt{UltraNest} \citep{Buchner2021Ultranest} for Bayesian Parameter Estimation. BXA finds the best fit, computes the associated error bars and marginal probability distributions, see \cite{Buchner2023arXiv230905705B} for a comprehensive tutorial on BXA. We run the BXA solver with default parameters, but note that there are different options to speed up BXA, including the possibility to parallelize BXA over multiple cores, as discussed in \cite{Buchner2023arXiv230905705B}.

We now introduce a method to restrict the prior range, with the objective of providing the network a training sample that is not too far from the targeted observation(s). This derives in part from the challenge, that for this work the generation of spectra, the inference, the generation of the posteriors should be performed on a MacBook Pro 2.9 GHz 6-Core Intel Core i9, within a reasonable amount of time.
\begin{table}[!ht]
\begin{center}
\begin{tabular}{lll}
    \hline
    \multicolumn{3}{c}{Model 1: \texttt{tbabs$*$powerlaw}}\\
    \hline
    \nh & $[0.1,0.3]$ & Uniform\\
    Gamma & $[0.5,2.5]$ & Uniform\\
    NormPL & $[0.01,100]$ & Uniform or Log Uniform \\
    \hline
    \multicolumn{3}{c}{Model 2: \texttt{tbabs$*$(powerlaw+bbodyrad) }}\\
    \hline
    \nh & $[0.15,0.35]$ & Uniform\\
    Gamma & $[1.,3.]$ & Uniform\\
    NormPL & $0.1,10.]$ & Uniform or Log Uniform\\
    kTbb & $[0.3,3.]$ & Uniform \\
    NormBB & $[100.,1000.]$ & Uniform or Log Uniform\\
    \hline
    \multicolumn{3}{c}{Model 3: \texttt{tbabs$*$(powerlaw+bbodyrad) }}\\
    \hline
    \nh & $[0.2]$ & Fixed\\ 
    Gamma & $[1.7]$ & Fixed\\ 
    NormPL & $0.1,10.]$ & Uniform or Log Uniform\\
    kTbb & $[0.1,4.5]$ & Uniform \\
    NormBB & $[0.1,100000.]$ & Uniform or Log Uniform\\
    \hline
\end{tabular}
\end{center}
\caption{Prior assumption for the emission models considered. The models are defined following the \xspec\ convention. \nh\ is given in units of equivalent hydrogen column (in units of $10^{22}$ atoms cm$^{-2}$). Gamma is the power law photon index. NormPL is the power law normalization at 1 keV in units of photons/keV/cm$^2$/s. kTbb is the blackbody temperature in keV. NormBB is the blackbody normalization given as $R_{\rm km}^2/D_{10}^2$, where $R_{\rm km}$ is the source radius in km and $D_{10}$ is the distance to the source in units of 10 kpc.}
\label{tab:simulation_set_ups}
\end{table}
 
\section{Generating an efficient training sample}
\label{restricted_prior}
The density of the training sample depends on the range of the priors and the number of simulations. As the training time increases with the size of the training sample, ideally, one would like to train the network with a limited number of simulated spectra that are not too far from the targeted observation(s), yet covering plainly the observation(s). Here, we consider two methods to restrict the priors: one in which we train a network to retain the “good” samples of $\Vec{\theta}_i$ matching a certain criterion (Sect. \ref{classifier}), and one in which we perform a coarse inference of the targeted observation(s) (Sect. \ref{coarse_inference}).

\subsection{Simulation set-up}

Let us first describe our simulation set-up. We assume a simple emission model consisting of an absorbed power law with three parameters: the column density (\nh), the photon index (Gamma, $\Gamma$) and the normalization of the power law at 1 keV (NormPL). We used the  \texttt{tbabs} model to take into account interstellar absorption, including the photoelectric cross-sections and the element abundances to the values provided by \cite{Verner_1996} and \cite{Wilms2000ApJ...542..914W}, respectively. In \xspec\ terminology, the model is \texttt{tbabs$*$Powerlaw}. For the simulations, we used NICER response files (for the observation identified in the NICER HEASARC archive as OBSID1050300108, see later). The simulated spectra are grouped in 5 consecutive channels so that each spectrum has $\sim 200$ bins covering the 0.3 and 10 keV range. The initial range of prior is given in Table \ref{tab:simulation_set_ups} for the model 1 set-up. We assume uniform priors in linear coordinates for \nh\ and for $\Gamma$ and in logarithmic coordinates for the power law normalization. The generation of synthetic spectra is done within \jaxspec\footnote{\jaxspec\ is a pure Python library for statistical inference on X-ray spectra. It allows to simply build spectral models by combining components, and fit it to one or multiple observed spectra. It builds on \texttt{JAX}  \citep{jax2018github} to enable computation on the GPU and just-in-time compilation. The differentiable framework enables advanced sampling algorithms such as \texttt{NUTS} \citep{hoffman2011nouturn}. The \texttt{v0.0.2} of \jaxspec\, which was used in this work, is available at \url{https://github.com/renecotyfanboy/jaxspec}, along with the \sixsa\ package at \url{https://github.com/dbxifu/SIXSA}. More details will be available in Dupourqué \& al. (in preparation).}, which offers parallelization of a fakeit-like command in \xspec\ (Dupourqué et al. in preparation). The generation of 10000 simulations takes about 10 seconds (including a few seconds of just-in-time compilation). In this first paper, we do not consider instrumental background. However, we note that if a proper analytical model exists for the background, the network could be trained to learn about the source and the background spectra simultaneously (with more free model parameters than in the source alone case). This would come at the expense of increasing the size of the training sample, hence inference time. The background spectrum could also be incorporated simply as a nuisance parameter, by adding to each bin of each simulated spectrum a number of counts with an empirical distribution estimated from the background spectrum. This would increase the dispersion in the spectrum, which would translate into an additional source of variance on the constraints of the model parameters. More details on this approach will be provided in the forthcoming paper by Dupourqué \& al. (in preparation).

\subsection{A restricted prior}
\label{classifier}
For the first method, we train a ResNet classifier \citep{he2015deep} to restrict the prior distributions \citep{Lueckmann2017arXiv171101861L, deistler2022pnas2207632119}. \texttt{sbi} can be used to learn regions of the parameter space producing valid simulations and distinguish them from regions that lead to invalid simulations. The process can be iterative, expecting that, as more simulations are fed to the classifier, the rejection rate will increase, and the restricted prior will shrink. It can be stopped when the fraction of valid simulations exceeds a given threshold, depending on the criterion. In all cases, it is recommended to perform sanity checks of the  coverage of the restricted prior for the observation(s) to fit. The user has to define a decision criterion for the classifier. For demonstrating the working principle of SBI-NPE under various statistical regimes (total number of counts in the X-ray spectra, see below), the first criterion we chose is to restrict the total number of counts in the spectra within a given range. Secondly, we have also considered a criterion such that the valid simulations are the ones providing the lowest \cstat\, computed from the observation to fit \citep{Cash1979ApJ...228..939C}. Once the simulations are produced, the classifier is also very fast (minute timescales), depending on the condition to match and the number of model parameters and size of the training sample (as an indication, for the three parameter model considered below, a few thousands simulations are required). We note that such a classifier is straightforward to implement and could be coupled to classical X-ray spectral fitting, to initialize the fit closer to the best fit solution, and thus reduce the likelihood of getting stuck into a local false minimum.

\begin{figure}[!h]
    \centering
\includegraphics[width=0.495\textwidth]{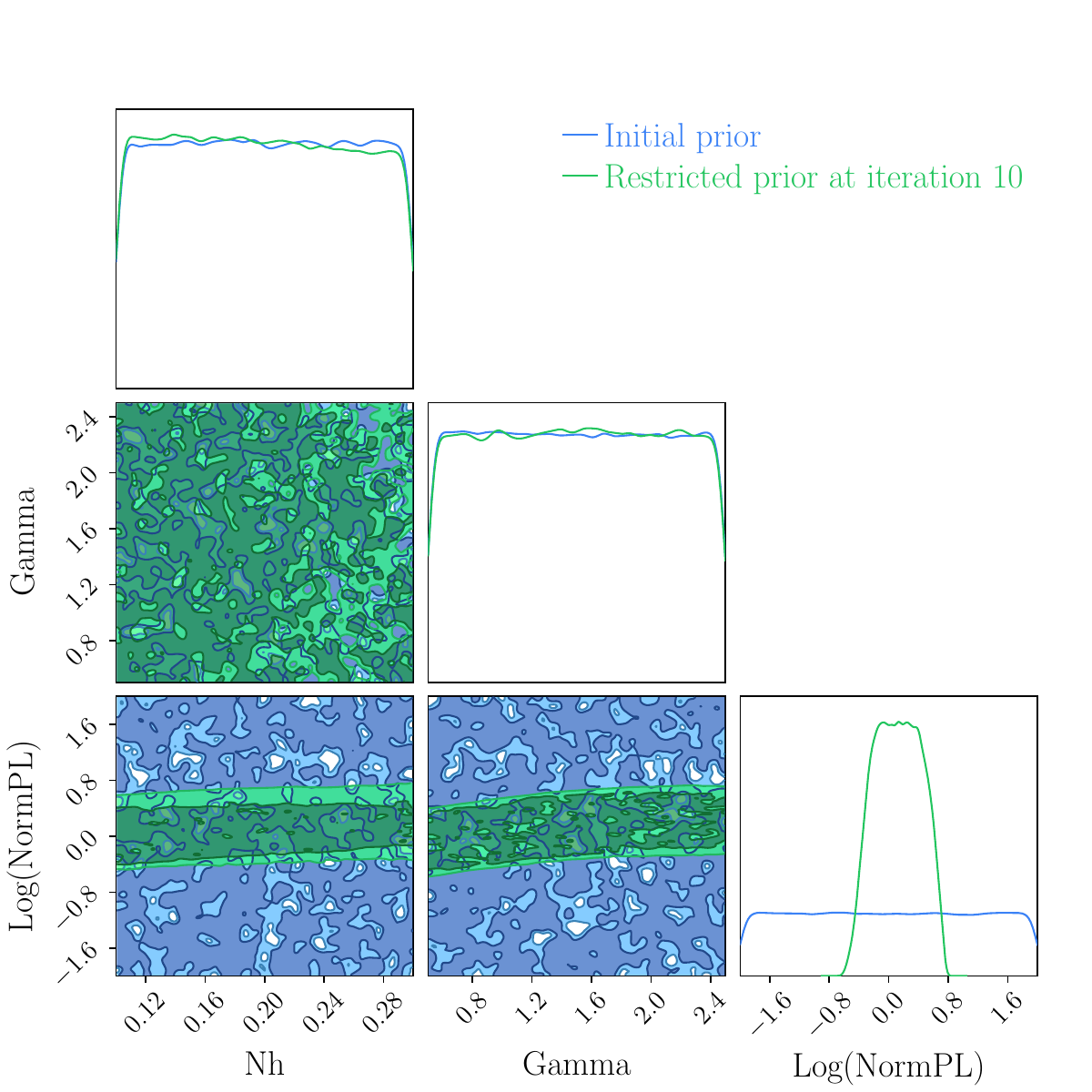}
    \caption{The initial and restricted prior, tuned to produce spectra that have between 10000 and 100000 counts for a \texttt{tbabs$*$powerlaw} model (at its tenth iteration). As expected, only a restricted range of the power law normalization can deliver the right number of counts.}
    \label{fig:db_sd_f2}
\end{figure}

The first statistical regime to be probe is the so-called high-count Gaussian regime. We define the integration time of the simulated spectra, identical to all spectra and defined such that a reference model (\nh=$0.2\times10^{22}$ cm$^{-2}$, Gamma=$1.7$ and NormPL=1) corresponds to a spectrum with about 20000 counts (over 200 bins). This provides the reference spectrum (referred as \specgauss). We define the criterion for the restrictor, such that the valid simulations are the ones which have between 10000 and 100000 counts, ensuring that the reference spectrum (\specgauss) is well covered. In Fig.~\ref{fig:db_sd_f2}, we show the initial and final round of the restricted prior, derived with the above criterion. Given the integration time for the spectra, only a restricted set of model parameters, mostly the normalization of the power law component, can deliver the right number of counts per spectrum.

\begin{figure}[!h]
    \centering
\includegraphics[width=0.495\textwidth]{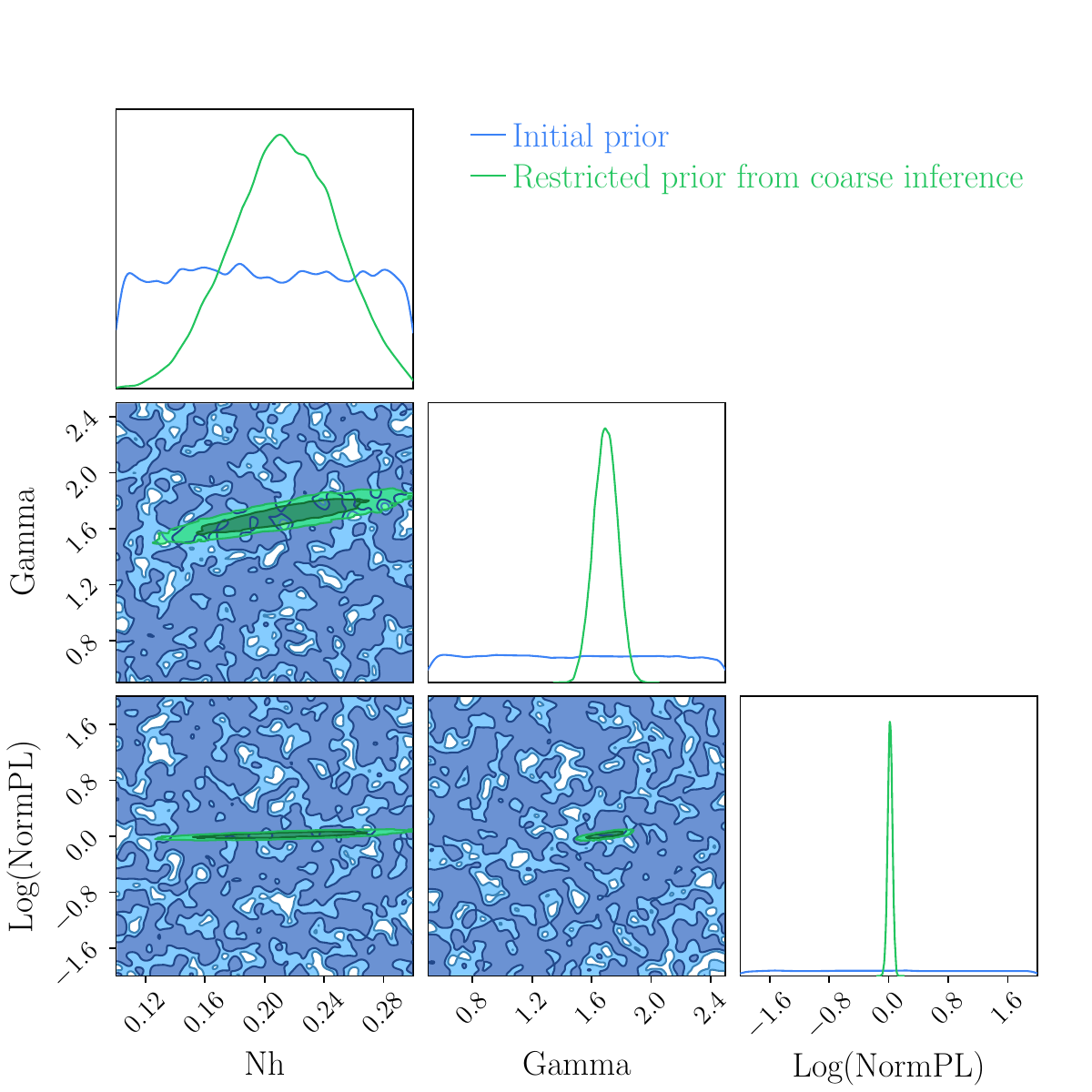}
    \caption{The initial and restricted prior computed from a coarse and quick inference of a reference spectrum of 20000 counts. Such a coarse inference can be seen as the first step of a multiple round inference.}
    \label{fig:db_sd_f3}
\end{figure}
\subsection{Coarse inference}
\label{coarse_inference}
\begin{figure}[!h]
    \centering
\includegraphics[width=0.495\textwidth]{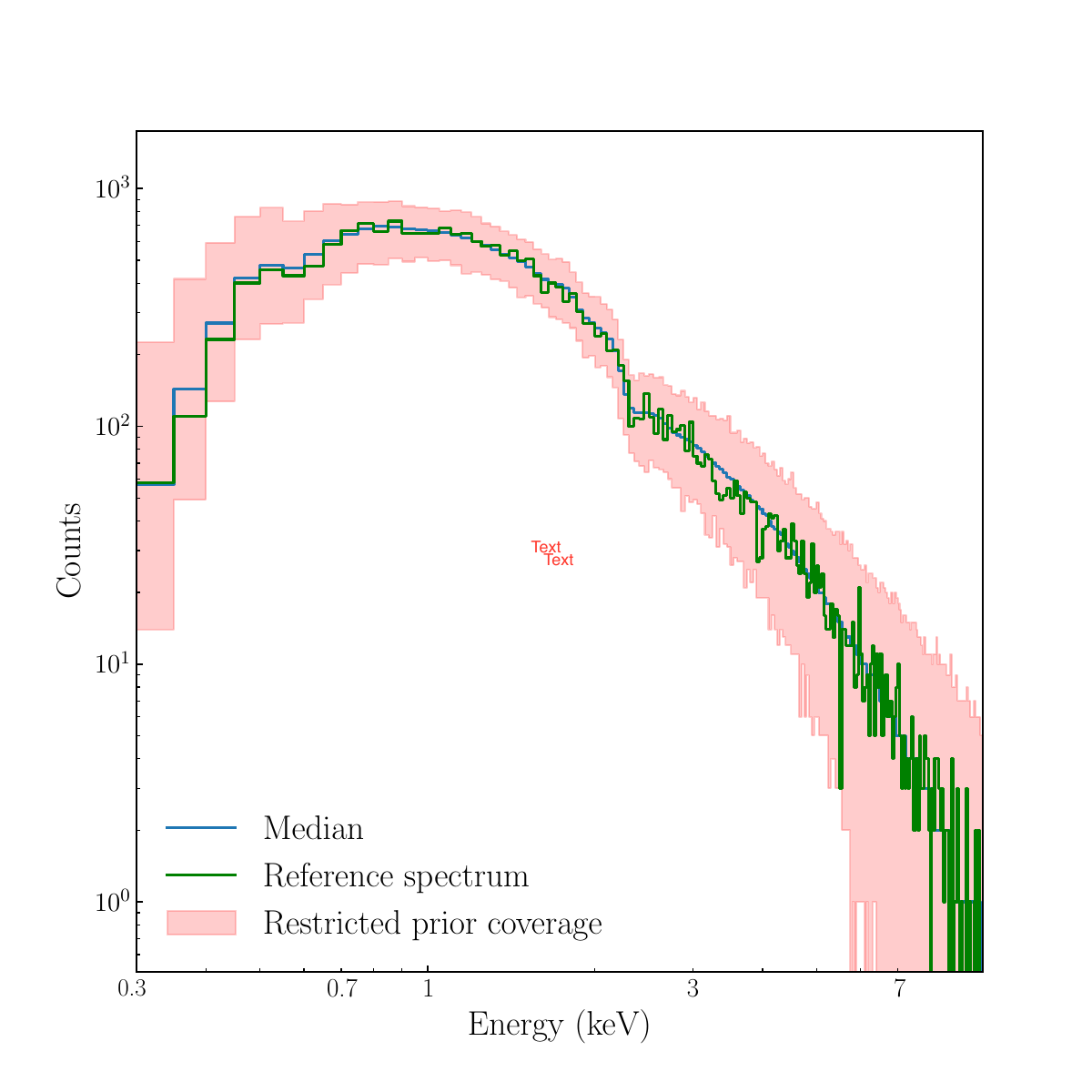}
    \caption{A prior predictive check showing that the coarse inference provides good coverage of the reference spectrum (\specgauss). The coverage of the restricted prior is indicated in red. The median of the spectra sampled from the restrictor (blue line) is to be compared with the reference spectrum (green line).}
    \label{fig:db_sd_f4}
\end{figure}

The second method uses a coarse inference, and can be considered as the first step of a multiple round inference. A coarse inference is when the network is trained with a limited number of samples (see Sec. \ref{gaussian_regime} for the parameters used to run the inference). The posterior conditioned at the reference observation is then used as the restricted prior. In Fig.~ \ref{fig:db_sd_f3}, we present the result of a coarse inference of the above reference spectrum (\specgauss). 5000 spectra are generated from the initial prior as defined in Tab.~\ref{tab:simulation_set_ups} for the above model, fed to the neural network, and the posterior distributions are computed at the reference spectrum. The training for such a limited sample of simulation, for three parameters, takes about 1 minute. As can be seen, the prior range is further constrained to narrower intervals. Generating a sample of 10000 spectra with parameters from this restricted prior shows that the reference spectrum is actually close to the median of the sample of the simulated spectra (see Fig.~\ref{fig:db_sd_f4}). The robustness of SBI-NPE against local false minima trapping (see Sect.~\ref{Sensitivity to local minima}), guarantees good coverage of the observation from the restricted prior.

\section{Single round inference}
\label{Single round inference}
Starting from the restricted prior, one can then draw samples of $\Vec{\theta}_i$ and generate spectra applying the Poisson count statistics in each spectral bin. The spectra are then binned the same way as the reference observation (grouped by 5 adjacent channels between 0.3 and 10 keV), and injected as such in the network (no zero mean scaling, no component reduction applied, see however Sect. \ref{Dimension reduction with the Principal Component Analysis}). For each run, we generate both a training and an independent test sample.

\subsection{In the Gaussian regime}
\label{gaussian_regime}
In the first case, we run the classifier with the criterion that each of the simulated spectra assuming an absorbed power law model having between 10000–100000 counts, spread over $\sim 200$ bins, are valid simulations. This covers the Gaussian regime. 

\begin{figure}[!t]
    \centering
    \includegraphics[width=0.465\textwidth]{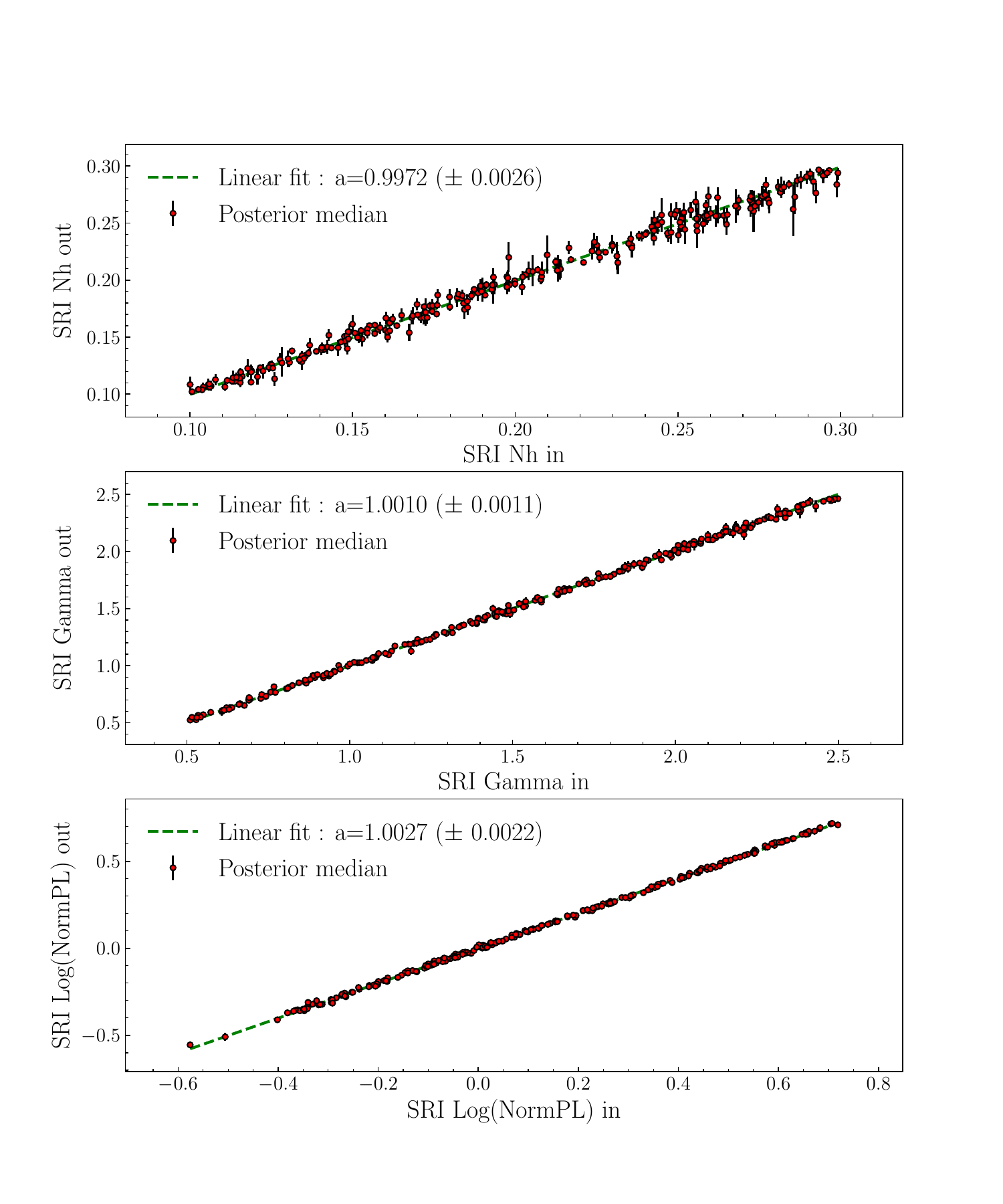}
    \caption{Inferred model parameters versus input model parameters for the case in which spectra have between 10000–100000 counts, spread over 200 bins. A single round inference is performed on an initial sample of 10000 simulated spectra. The median of the posterior distributions is computed from 20000 samples and the error on the median is computed from the 68\% quantile of the distribution. The linear regression coefficient is computed for each parameter over the 500 test samples.}
    \label{fig:db_sd_f5}
\end{figure}

For three parameters, we generate a set of 10000 spectra. The network is trained in $\sim 3$ minutes. The inference is performed with the default parameters of \texttt{SNPE\textunderscore C} as implemented in \texttt{sbi}, which is using 5 consecutive MADEs with 50 hidden states each for the density estimation. It then takes about the same time to draw 20000 posterior samples for 500 test spectra, i.e. to fit 500 different spectra with the same network. The inferred model parameters versus the input parameters are shown in Fig.~ \ref{fig:db_sd_f5}. As can be seen, there is an excellent match between the input and output parameters: the linear regression coefficient is very close to 1. For \nh, a minimal bias is observed towards the edge of the sampled parameter space. 

\begin{figure}
    \centering
    \includegraphics[width=0.465\textwidth]{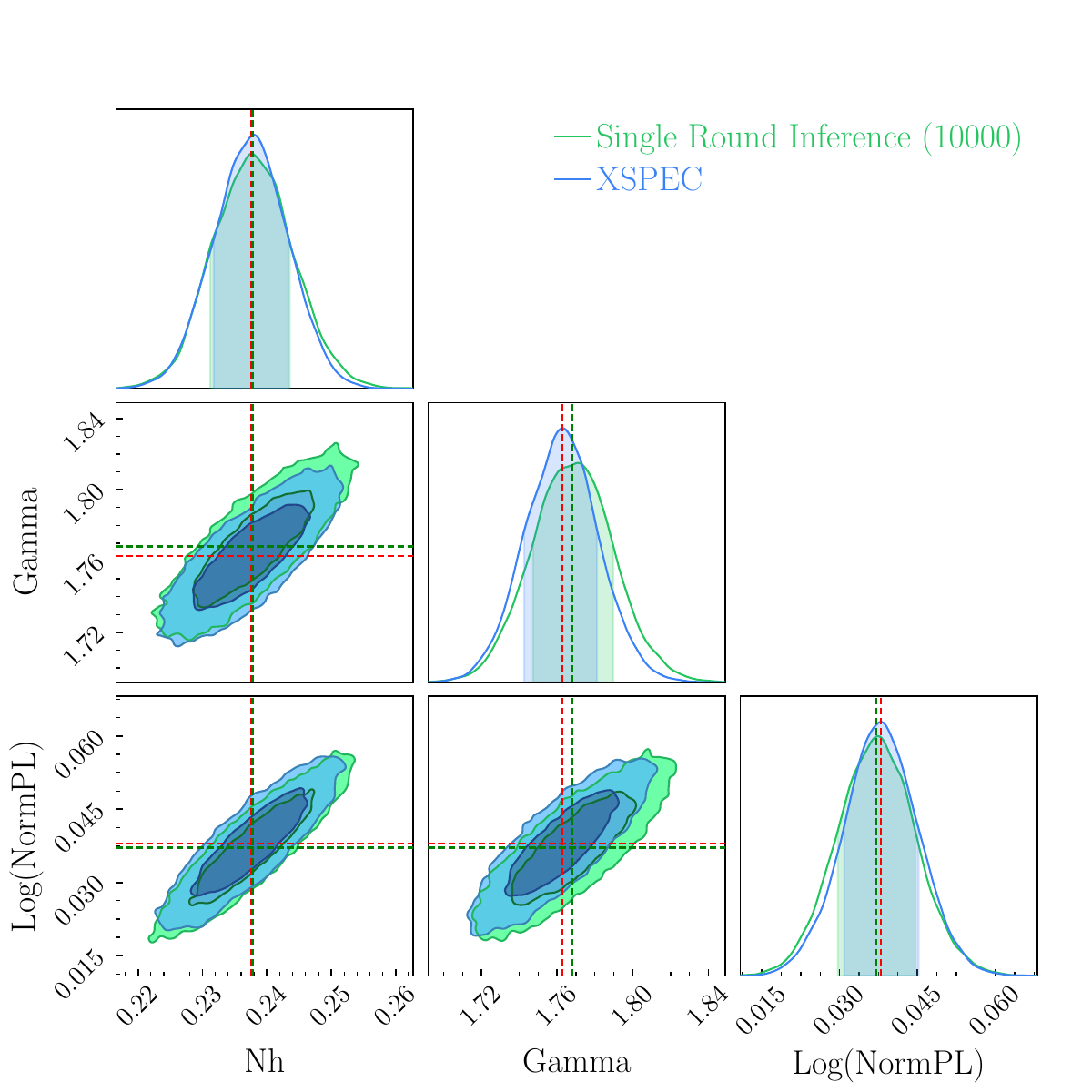}
    \caption{The posterior distribution estimated for the reference absorbed power law spectrum (\specgauss) as inferred from a single round inference with a network trained on 10000 samples (green). The posterior distribution inferred from a Bayesian fit with \xspec\ is also shown in blue.}
    \label{fig:db_sd_f6}
\end{figure}

We generate the posterior distributions for the reference absorbed power law spectrum and compare it with the posterior distribution obtained from \xspec\, switching on Bayesian inference \footnote{\xspec\ by default uses uniform priors for all parameters. For this run, we considered Jeffrey's prior for the normalization of the power law instead of a log uniform distribution.  \xspec\ adds to the fit statistic a contribution from the prior as $-2\ln{P_{prior}}$ \citep{Arnaud1996ASPC..101...17A} (which is in our case a positive contribution equal to the $2\ln$(NormPL)). Such a contribution is removed when comparing the results of \xspec\ run with Bayesian inference, with the results of SBI-NPE. We note that in BXA, Jeffrey's priors have been deprecated for log uniform priors.}. The comparison is shown in Fig.~ \ref{fig:db_sd_f6}. There is an excellent match between the best fit parameters, but also the posterior distribution, including their spread. In both cases, the \cstat\ of the best fit is within $1\sigma$ of the expected \cstat\ following \cite{Kaastra2017A&A...605A..51K}. Note that running \xspec\ to derive the posterior distribution takes about the same time as training the network. 

In the above case, we compute the goodness of the fit by comparing the measured fit statistic with its expected value following \cite{Kaastra2017A&A...605A..51K}. \xspec\ offers the possibility to perform a Monte Carlo calculation of the goodness-of-fit (\texttt{goodness} command), applicable in the case when the only source of variance in the data is counting statistics. This command simulates a number of spectra based on the model and returns the percentage of these simulations with the test statistic less than that for the data. If the observed spectrum is produced by the model then this number should be around 50\%. \xspec\ can generate simulated spectra either from the best fit parameters or from a Gaussian distribution centered on the best fit, with sigma computed from the diagonal elements of the covariance matrix. \xspec\ can either fit each simulated spectra or return the fit statistic calculated immediately after creating the simulated spectrum. Fitting the data is the option by default, but at the expense of increasing the run time. Such a feature would be straightforward to implemented with SBI-NPE and amortized inference. The set of parameters of the simulated spectra can be drawn from the posterior distribution conditioned at the observation, and the posterior distributions for each simulated spectrum can be computed from the already trained amortized network (the fit statistic associated with each simulated spectrum is as usual derived from the median of the so-computed posterior distributions). With the simulation set-up considered here, generating the simulated spectra and the posterior distributions is very fast (few minutes for 1000 spectra).

\begin{figure}
    \centering
    \includegraphics[width=0.465\textwidth]{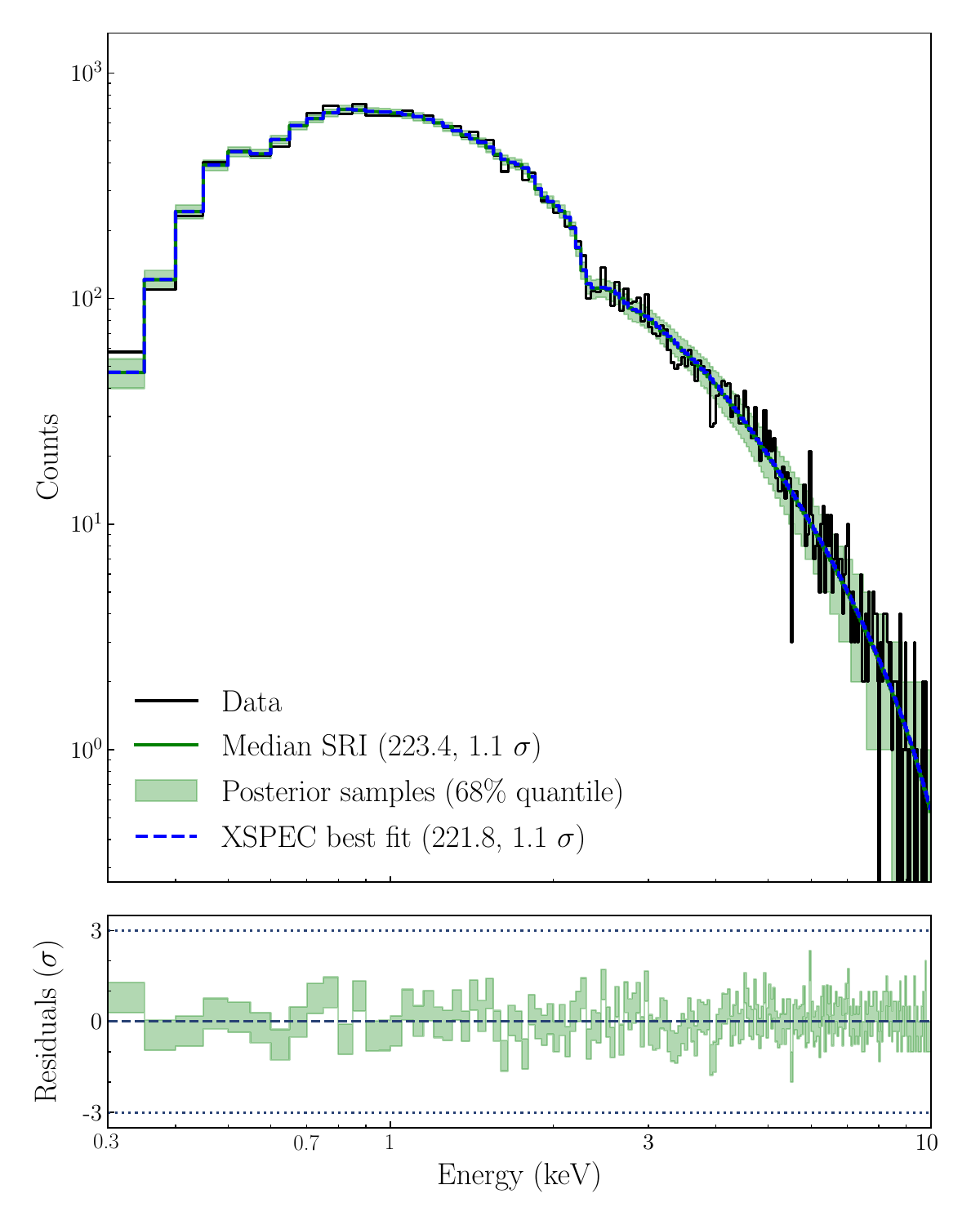}
    \caption{Top) The count spectrum corresponding to the reference absorbed power law spectrum (\specgauss), together with the folded best fit model from both SBI-NPE (green solid line) and \xspec\ (blue dashed line). Bottom) The residuals of the best fit from SBI-NPE.}
    \label{fig:db_sd_f7}
\end{figure}

We now show the count spectrum corresponding to the reference spectrum of this run, together with the folded model and the associated residuals, in Fig.~\ref{fig:db_sd_f7} for both SBI-NPE and \xspec. As expected from Fig.~ \ref{fig:db_sd_f6}, there is an excellent agreement between the fits with the two methods. 

\begin{figure}[!h]
    \centering
    \includegraphics[width=0.485\textwidth]{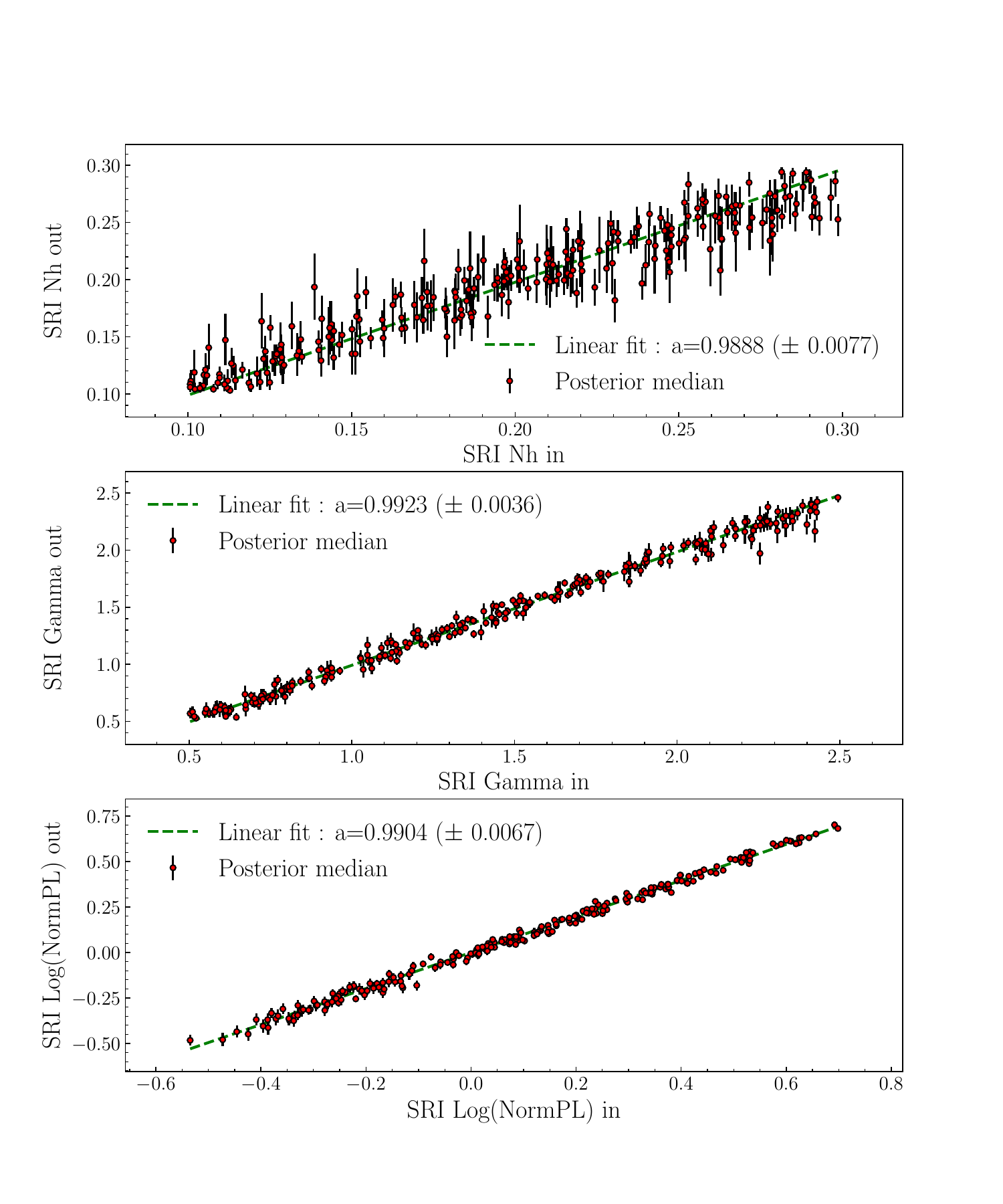}
    \caption{Inferred model parameters versus input model parameters for the case in which spectra have between 1000–10000 counts, spread over 200 bins. The neural network is trained with 20000 simulations and the posteriors for 500 test spectra are then computed. The medians of the posteriors are computed from 20000 samples and the error on the median is computed from the 68\% quantile of the distribution. The linear regression coefficient is computed for each parameter over the 500 test samples.}
        \label{fig:db_sd_f8}
\end{figure}
\subsection{In the Poisson regime}
The above results can be considered as encouraging, but to test the robustness of the technique, we must probe the low count Poisson regime. We repeat the run above but this time generating spectra from a restricted prior generating 200 bin spectra with a number of counts ranging between 1000–10000 for the absorbed power law model. The integration time of the spectra is scaled down from the Gaussian case, so that the reference model (\nh=$0.2\times10^{22}$ cm$^{-2}$, Gamma=$1.7$ and NormPL=1) now corresponds to a spectrum of $\sim 2000 $ counts (referred as \specpoiss). To account for the lower statistics, we train the network with a sample of 20000 spectra, instead of 10000 for the case above. The training time still takes about 3 minutes. We generate the posterior for a test set of 500 spectra, and this takes again about 3 minutes. Similar to Fig.~ \ref{fig:db_sd_f5}, we show the input and inferred model parameters for the test sample in Fig.~\ref{fig:db_sd_f8}. As in the previous case, although with larger error bars, accounting for the lower statistics of the spectra, there is an excellent match between the two quantities, the linear regression coefficient remains close to 1, with some evidence for the small bias on \nh\ at both ends of the parameter range increased. 

\begin{figure}
    \centering
    \includegraphics[width=0.485\textwidth]{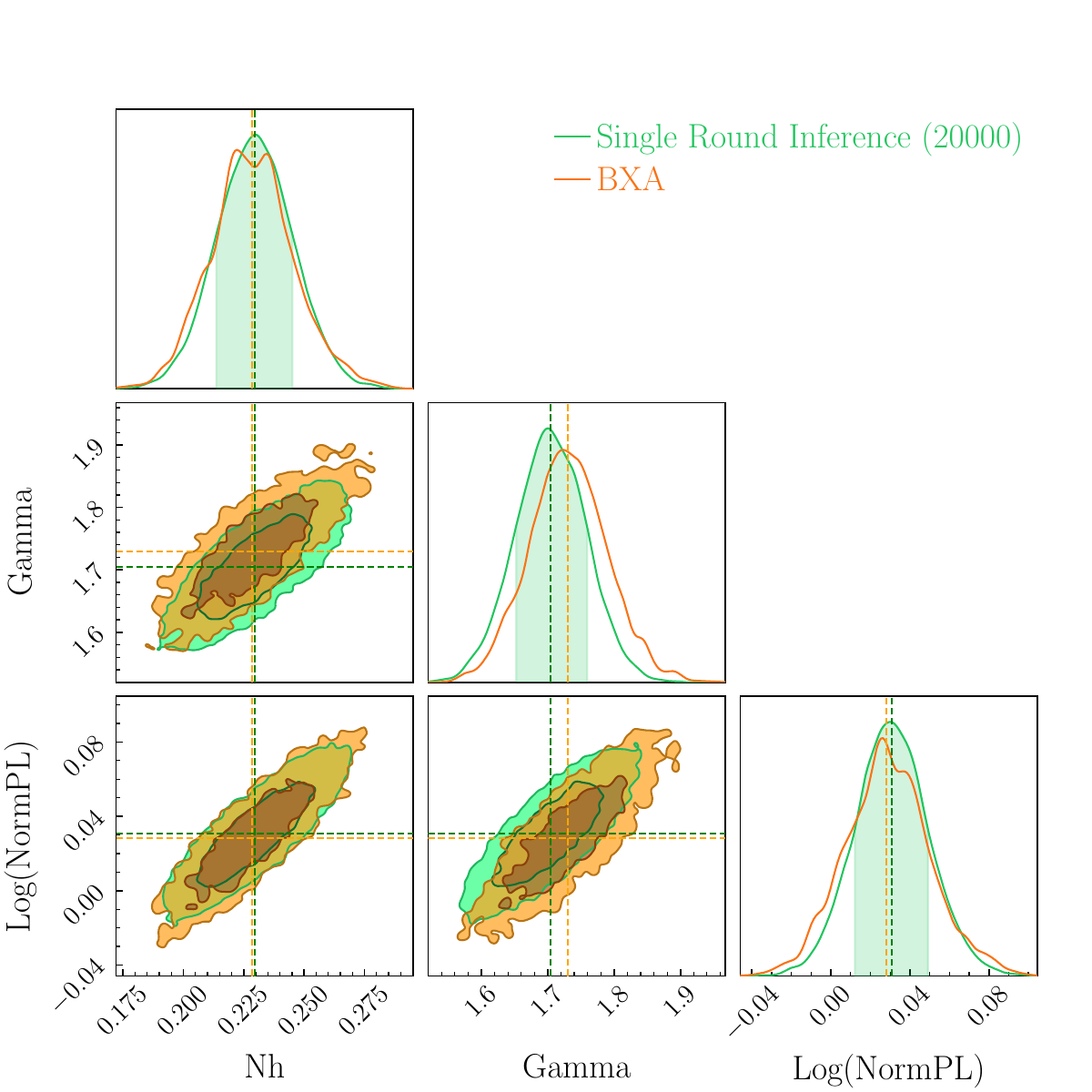}
    \caption{The posterior distributions for a reference spectrum of 2000 counts as derived from SBI-NPE with a single round inference of a network trained with 20000 samples (green) and by BXA in orange.}
    \label{fig:db_sd_f9}
\end{figure}

We generate the Posterior distribution for the reference spectrum (\specpoiss), which we also fit with BXA (assuming the same priors as listed in Tab.~\ref{tab:simulation_set_ups} for the model 1 set-up). The posterior distributions are compared in Fig.~\ref{fig:db_sd_f9}, showing again an excellent agreement. Not only the best fit parameters are consistent with one another, but as important, the widths of the posterior distribution are also comparable. This demonstrates that SBI-NPE generates healthy posteriors. Note that the time to run BXA (with the default solver parameters and without parallelization) on such a spectrum is comparable with the training time of the network.
\section{Multiple round inference} 
\label{Multiple round inference}
\begin{figure}
    \centering
    \includegraphics[width=0.485\textwidth]{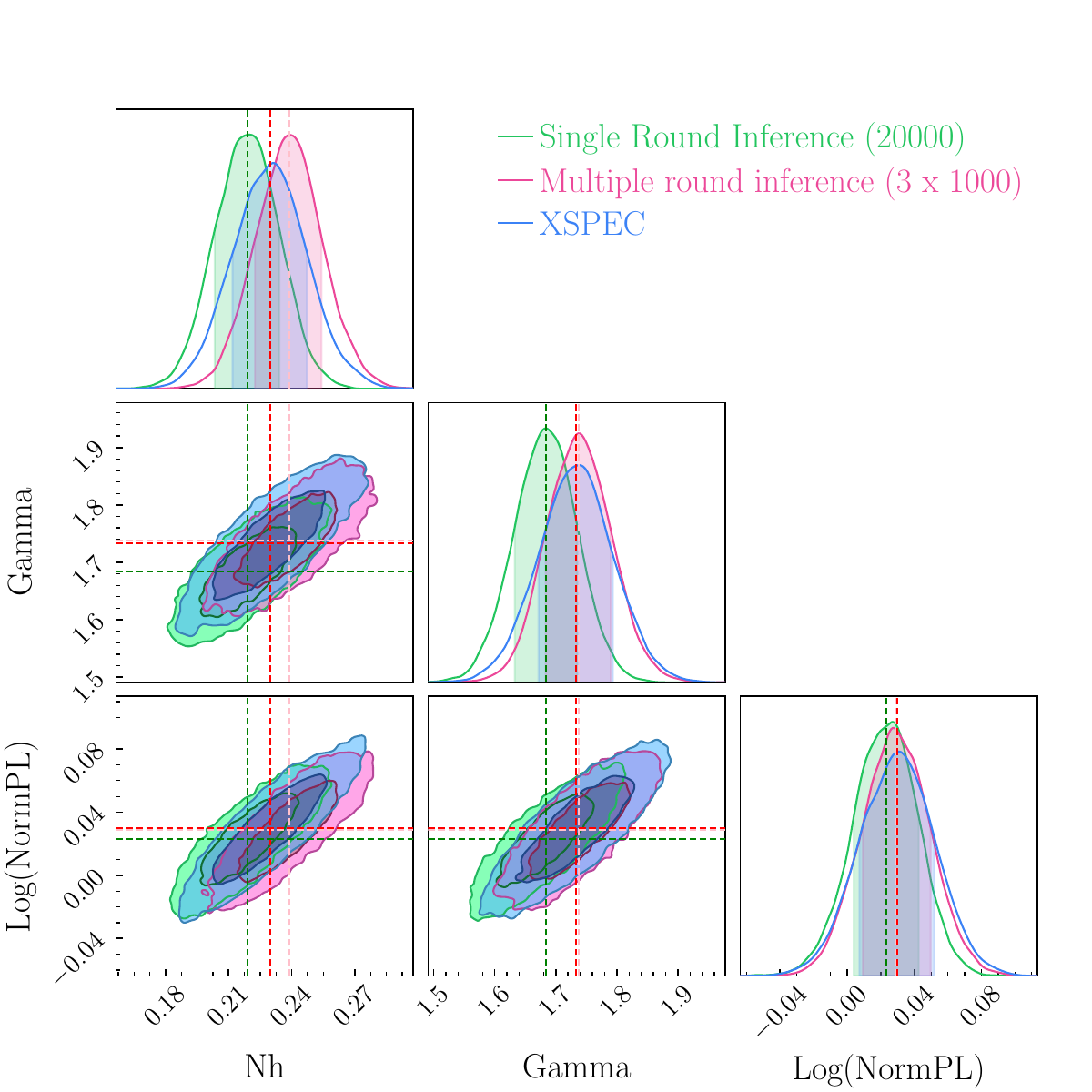}

    \caption{The posterior distributions for a reference spectrum of 2000 counts as derived from SBI-NPE with a single round inference of a network trained with 20000 samples (green), multiple round inference with three iterations, in which the network in trained with a sample of 1000 simulations (pink) and by \xspec\ (blue).}
    \label{fig:db_sd_f10}
\end{figure}

In the previous cases, the posterior is inferred using single-round inference. We are now considering multi-round inference, tuned for a specific observation; the reference spectrum of 2000 counts, spread over 200 bins (\specpoiss). For the first iteration, from the restricted prior, we generate 1000 simulations, and train the network to estimate the posterior distribution. In each new round of inference, samples from the obtained posterior distribution conditioned at the observation (instead of from the prior) are used to simulate a new training set used for training the network again. This process can be repeated an arbitrary number of times. Here we stop after three iterations. The whole procedure takes about 1.5 minutes. In Fig.~\ref{fig:db_sd_f10}, we show that multi-round inference returns best fit parameters and posterior distributions consistent with single-round inference (from a larger training sample) and \xspec. We thus confirm that multi-round inference can be more efficient than single round inference in the number of simulations and is faster in inference time. Its drawback is however that the inference is no longer  amortized (i.e. it will only apply for a specific observation).

\section{Sensitivity to local minima} 
\label{Sensitivity to local minima}
\begin{figure}
    \centering
    \includegraphics[width=0.485\textwidth]{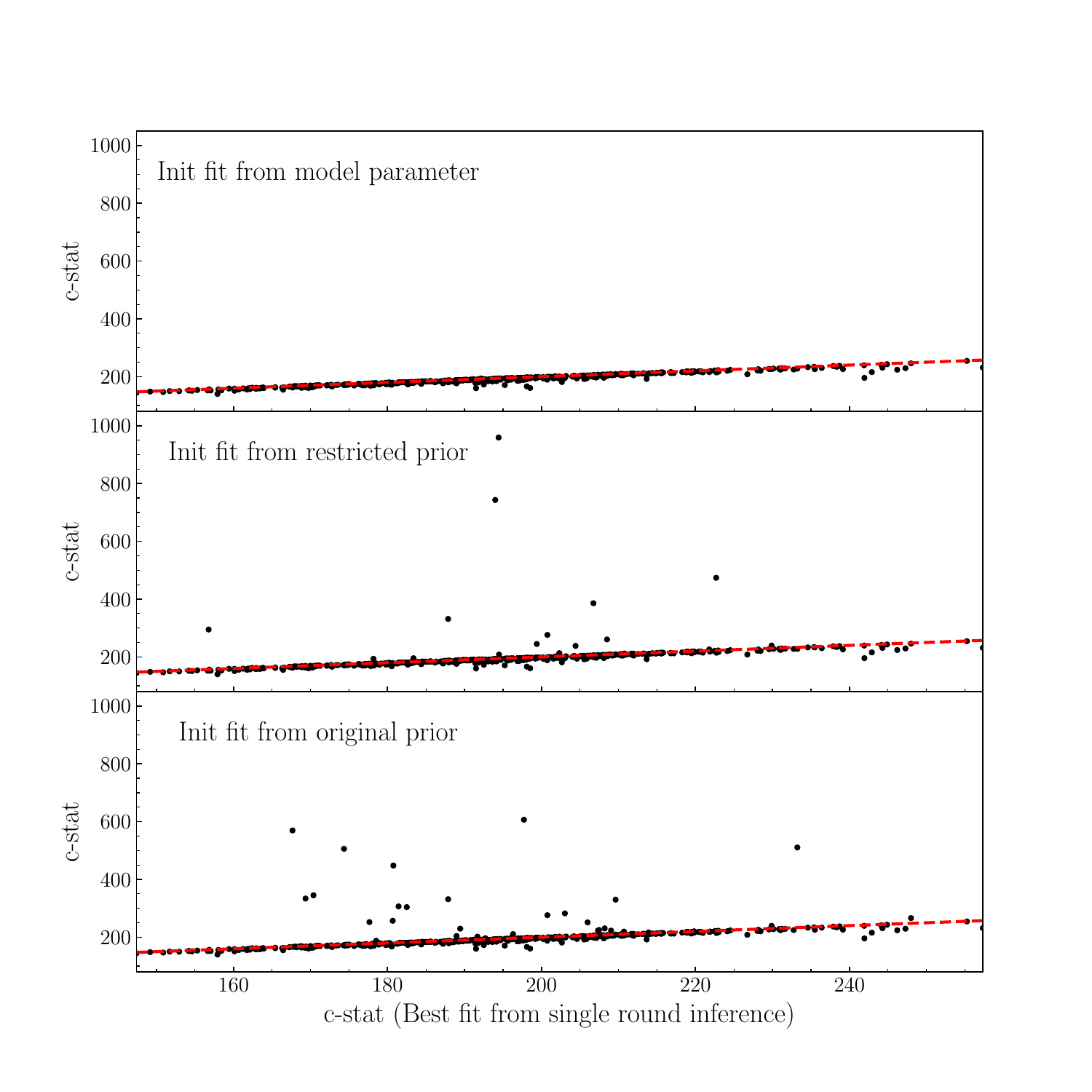}
    \caption{Comparison between \xspec\ spectral fitting (Fit statistic minimization) and SBI-NPE. The initial parameters of the \xspec\ fits are the input model parameter (top panel), a set of parameters generated from the restricted prior (middle panel), and a set of parameters generated from the initial prior (bottom panel). The y-scale is the same for the three panels. Outliers away from the red line are due to \xspec\ getting trapped in a local false minimum.}
    \label{fig:db_sd_f11}
\end{figure}

Fit statistic minimization algorithms may get stuck into local false minima. There are different workarounds, such as computing the errors on the model parameters to explore a wider parameter space, shaking the fits with different sets of initial parameters, using Bayesian inference, all at the expense of increasing the processing times. This is probably the reason why fit statistic minimization remains widely used, despite its known limitations. This makes worth the comparison of the sensitivity to local minima of SBI-NPE with \xspec\ is its common use. For this purpose, we are now going to consider a 5 parameter model combining two overlapping components, a power law and a blackbody. In the \xspec\ terminology, the model would be \texttt{tbabs$*$(powerlaw+blackbody)} (see Tab.~\ref{tab:simulation_set_ups} for the model 2 simulation set-up for the priors). We build a restricted prior so that such a model produces at least 10000 counts per spectrum, as decent statistics is required to constrain a 5 model parameters. We train a network with 100000 simulated spectra. We then generate the posterior of 500 spectra that we fit with \xspec, with three sets of initial parameters: the model parameters, a set of model parameters generated from the restricted prior, a set of model parameters from the initial prior (that do not meet necessarily our requirement on counts). We switch off Bayesian inference in the \xspec\ fits, and return for each of the fits, the best fit \cstat\ statistics. In Fig.~\ref{fig:db_sd_f11}, we compare the \cstat\ of SBI-NPE as derived from a single round inference with the \xspec\ fitting. This figure shows that SBI-NPE does not produce outliers, while the minimization does, at the level of a few percent. The latter is a known fact. The use of a restricted prior helps in reducing the trapping in local false minima, compared to considering the wider original prior, because the \xspec\ fits starts closer to the best fit parameters. The most favorable, and unrealistic, situation for \xspec\ is when the fit starts from the model parameters. In some cases, SBI-NPE produces minimum \cstat\ that are slightly larger than those derived from \xspec, indicating that the best fit solution was not reached. This may simply call for enlarging the training sample of the network for such a 5 parameter model, or considering multiple-round inference. 

\section{Dimension reduction with the Principal Component Analysis}
\label{Dimension reduction with the Principal Component Analysis}
\begin{figure}[!h]
    \centering
    \includegraphics[width=0.485\textwidth]{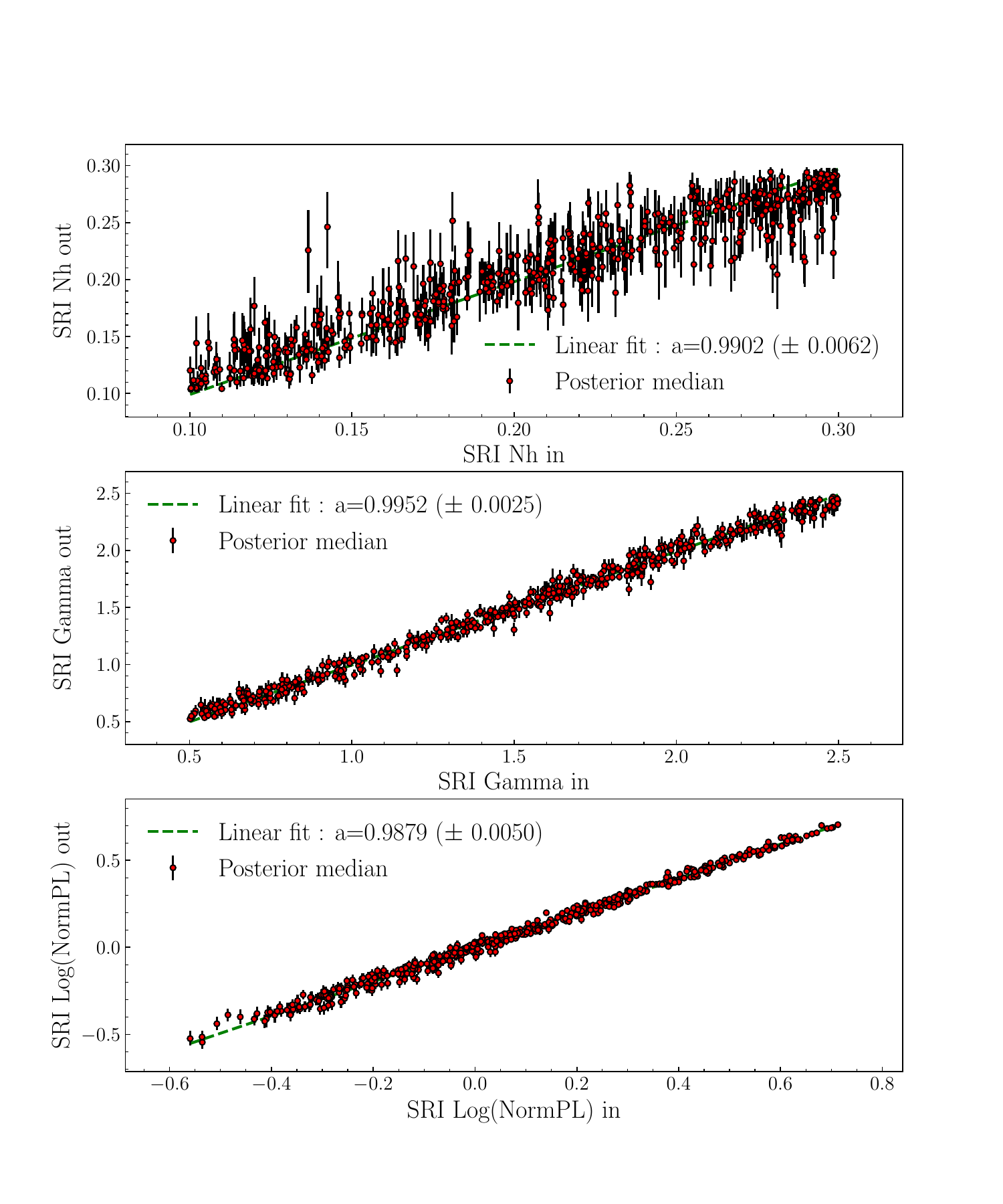}
        \caption{Inferred model parameters versus input model parameters for the case in which spectra have between 1000–10000 counts, spread over 200 bins. The best fit parameters are derived from a single round inference of a network trained by 20000 samples, reduced by the PCA, so that only 90\% of the variance in the samples is kept.}
    \label{fig:db_sd_f12}
\end{figure}

\begin{figure}[!h]
    \centering
    \includegraphics[width=0.485\textwidth]{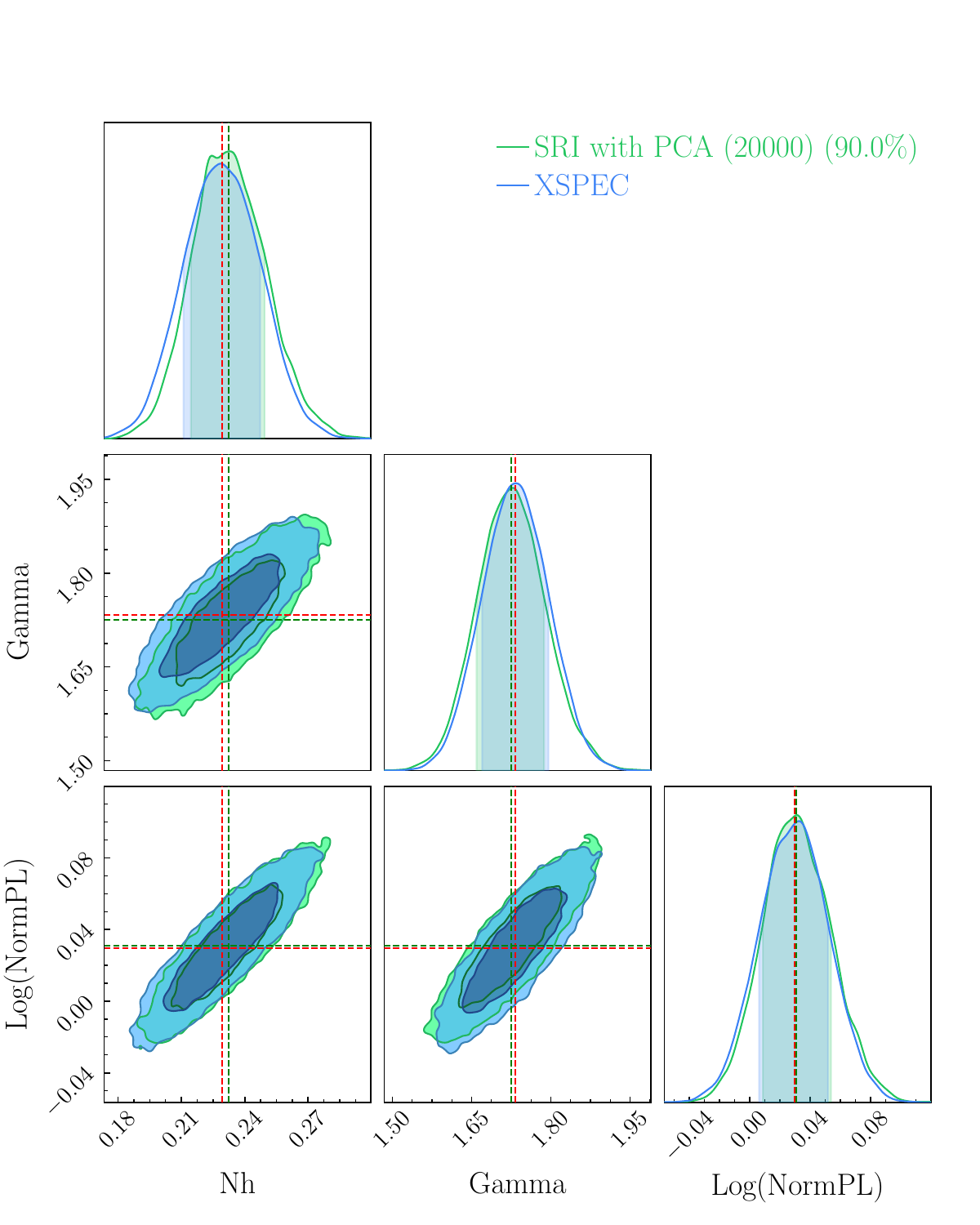}
        \caption{The posterior distributions for a reference spectrum of 2000 counts as derived from SBI-NPE with a single round inference of a network trained with 20000 samples (green) decomposed by the PCA (green, dimension reduced by a factor of 3) and \xspec\ (blue).}
    \label{fig:db_sd_f13}
\end{figure}

\cite{Parker2022MNRAS.514.4061P} introduced the use of principal component analysis (PCA) to reduce the dimension of the data to feed the neural network, and showed that it increased the accuracy of the parameter estimation, without any penalty on computational time, yet enabling simpler network architecture to be used. The PCA performs a linear dimension reduction using singular value decomposition of the data to project it to a lower dimensional space. Unlike \cite{Parker2022MNRAS.514.4061P}, we have here access to the posteriors and it is worth investigating whether such a PCA decomposition affects the uncertainty on the parameters estimates. Considering the run presented in Sect.~\ref{Single round inference} for the case of single round inference in the Poisson regime, we decompose the 20000 spectra with the PCA, as to keep 90\% of their variance (before that we scale the spectra to have a mean of zero and standard deviation of 1). This allows to reduce the dimension of the data from $20000 \times 200 $ to $20000 \times 60 $, i.e. a factor of 3 reduction, leading to a gain in inference time by a factor of 2. We show in Fig.~\ref{fig:db_sd_f12} the input and output parameters from a single round inference trained on dimension reduced data. As can be seen, there is still an excellent agreement between the two with the linear regression coefficient close to 1, although the bias on \nh\ at the edge of the prior interval seems to be more pronounced (the slopes for all parameters are less than 1, indicating that a small bias may have been introduced through the PCA decomposition). 
We show the posteriors of the fit of the reference spectrum of 2000 counts, in comparison with \xspec\ in Fig.~ \ref{fig:db_sd_f13} to show that the posteriors are not broadened by the dimension reduction. 
\section{Application to real data}
\label{Application to real data}
Having shown the power of the technique on mock simulated data, it now remains to demonstrate its applicability to real data, recorded by an instrument observing a celestial source of X-rays (and not data generated by the same simulator used to train the network). This is a crucial step in machine learning applications. 

We have considered NICER response files for the above simulations because we are now going to apply the technique to real NICER data recorded from \4u \citep{Gendreau2012SPIE.8443E..13G, Keek2018ApJ...856L..37K}. For the scope of the paper, we are going to consider two cases: a spectrum for the persistent X-ray emission (number of counts $\sim 200000$) and spectra recorded over a type I X-ray burst, when the X-ray emission shows extreme time and spectral variability. 
\subsection{NICER data analysis}
We have retrieved the archival data of \4u\ from HEASARC for the observation identifier (1050300108), and processed them with standard filtering criteria with the \texttt{nicerl2} script provided as part of  the \texttt{HEASOFTV6.31.1} software suite, as recommended from the NICER data analysis web page (NICER software version : \texttt{{NICER\_2022-12-16\_V010a}}). Similarly, the latest calibration files of the instrument are used throughout this paper (reference from the \texttt{CALDB} database is \texttt{xti20221001}). A light curve was produced between 0.3 and 7\,keV band, with a time resolution of 120\,ms, so that the type I X-ray burst could be located precisely. 

\subsection{Spectrum of the persistent emission} 
\begin{figure*}
    \centering
    \includegraphics[width=0.485\textwidth]{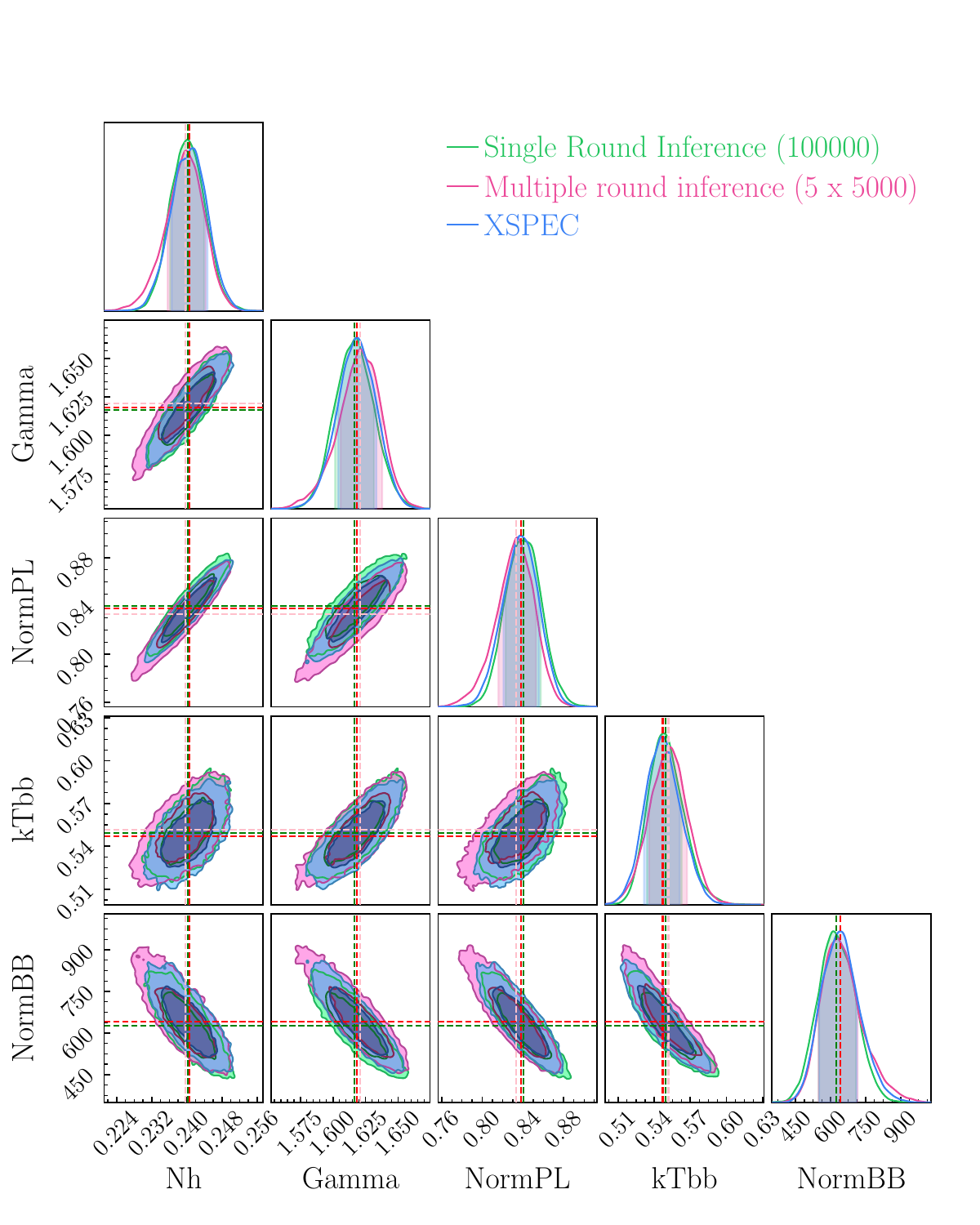}
    \includegraphics[width=0.45\textwidth]{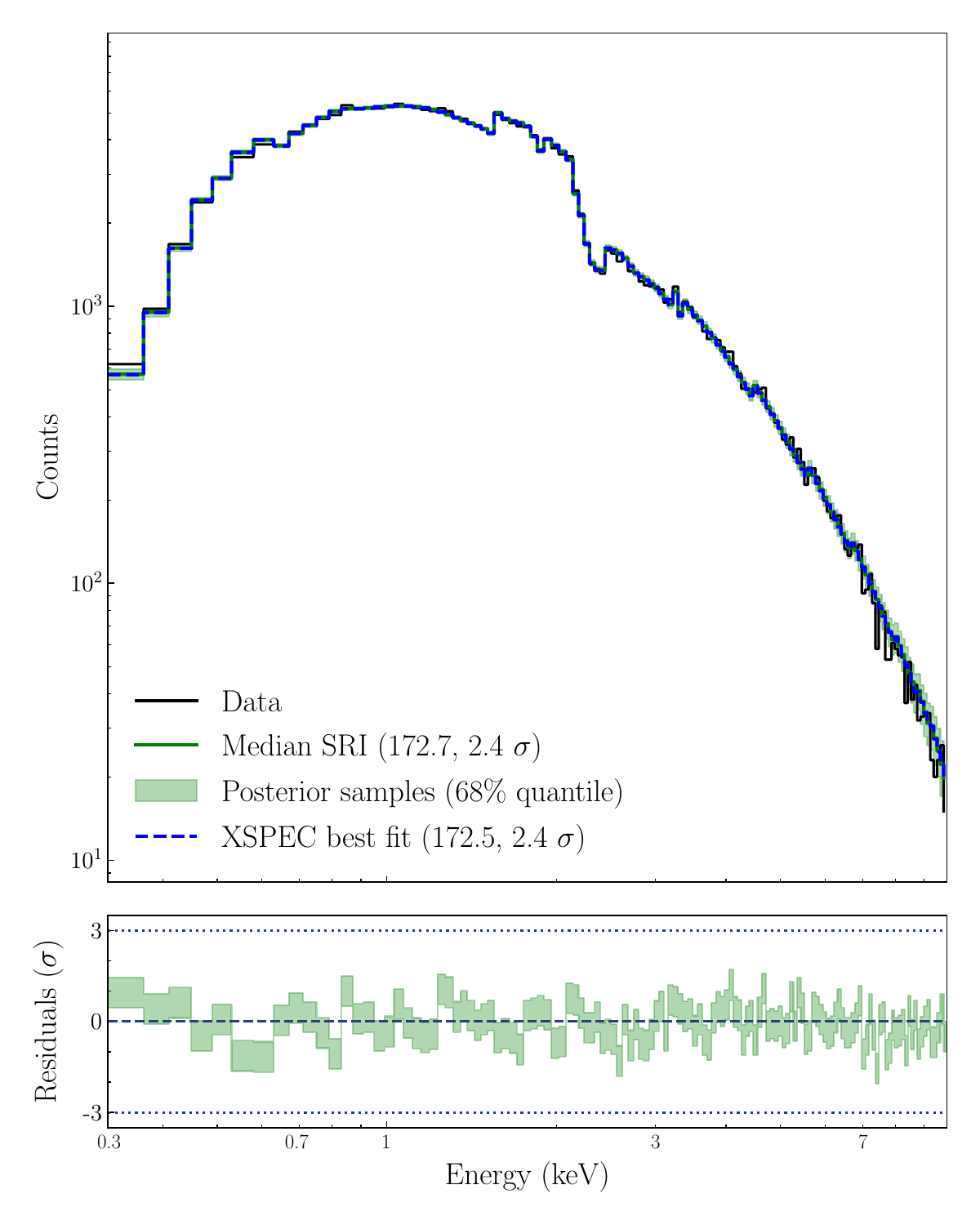}
    \caption{Left) Posterior distribution comparison between \xspec\ spectral fitting and SBI-NPE, as applied to the persistent emission spectrum of \4u (pre-burst). The spectrum is modeled as \texttt{tbabs$*$(blackbody+powerlaw)}. There is a perfect match between the three methods, not only on the best fit parameters but also on the width of the posterior distributions. Right) The count spectrum of the persistent emission, together with the folded model derived from both SBI-NPE and \xspec. The \cstat\ of the best fit is indicated together with its deviation against the expected value. }
\label{fig:db_sd_f14}

\end{figure*}

Once the burst time was located, we first extract a spectrum of the persistent emission for 200 seconds, ending 10 seconds before the burst. The spectrum is then modelled by a 5-component model, as the sum of an absorbed and a power law. In \xspec\ terminology, the model is \texttt{tbabs$*$(blackbody+powerlaw)}.  The initial range of prior is given in Tab.~\ref{tab:simulation_set_ups} for the model 2 set-up. For both SBI-NPE and \xspec\ spectral fitting, for a change, we consider uniform priors in linear coordinates for all the parameters. Similarly, for this observation, we build a restricted prior with the criterion that the classifier keeps 25\% of the model parameters associated with the lowest \cstat (considering a set of 5000 simulations for 5 parameters). From this restricted prior, we generate a rather conservative set of 100000 spectra for a single round inference, and a set of 5000 spectra for a multiple-round inference considering only three iterations. It takes about 40 minutes to train the network with 100000 spectra with 5 parameters, and 12 minutes for the 3 iteration multiple-round inference. The posterior distribution from single and multiple round inference, and the \xspec\ fitting are shown in Fig.~ \ref{fig:db_sd_f14}. As can be seen, there is a perfect match between \xspec\ and SBI-NPE, demonstrating that the method is also applicable to real data. We have verified that changing the assumptions for the priors, e.g. uniform in logarithmic scale for the normalizations of both the blackbody and the power law, yielded fully consistent results, in terms of best fit parameters, \cstat\ and posterior distributions. The same applies when using BXA instead of \xspec. This is the first demonstration to date that SBI-NPE performs equally well as state-of-the-art X-ray fitting techniques, on real data.

\subsection{The burst emission}
The first burst observed with NICER was reported by  \citep{Keek2018ApJ...856L..37K,Strohmayer2019ApJ...878L..27S,Yu2023arXiv231216420Y}. The burst emission is fitted with a blackbody model and a component accounting for some underlying emission, which in our case is assumed to be a simple power law. The blackbody  temperature and its normalization vary strongly along the burst itself, in particular in this burst, which showed the so-called photospheric expansion, meaning that the temperature of the blackbody drops while its normalization increases, to raise again towards the end of the burst. To follow spectral evolution along the burst, we extract fixed duration spectra (0.25 seconds), still grouped in 5 adjacent channels, so that all the spectra have the same number of bins (200). The number of counts per spectrum ranges from 400 to 5000, hence offering the capability of exploring the technique with real data in the Poisson regime. In this range of statistics, we cannot constrain a 5-parameter model. Hence, we fix the column density and the power law index to \nh=$0.2\times 10^{22}$ cm$^{-2}$ and $\Gamma=1.7$ respectively. The initial range of prior is given in Table \ref{tab:simulation_set_ups} for the model 3 set-up.

\begin{figure*}[!h]
    \centering
    \includegraphics[width=0.54\textwidth]{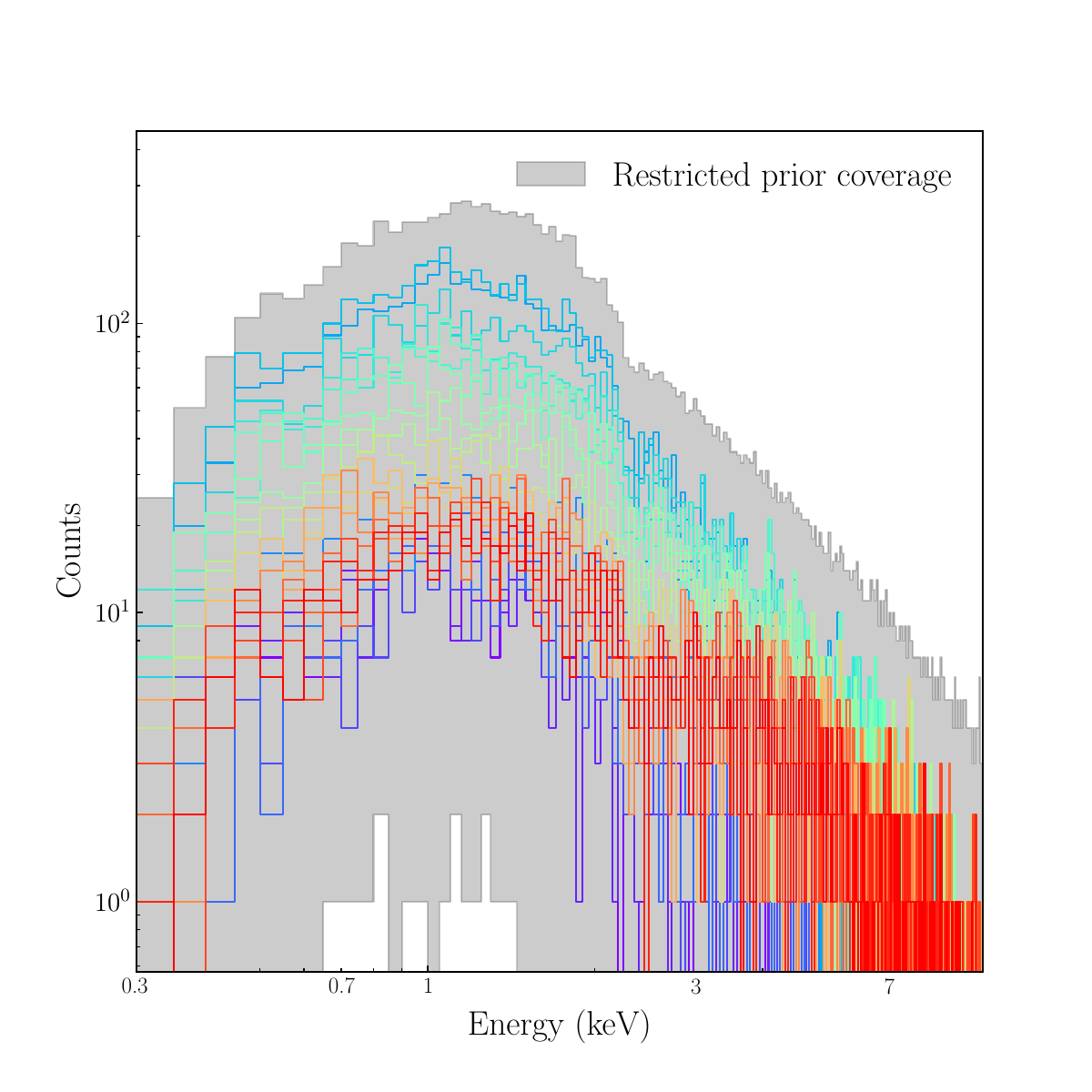}
    \includegraphics[width=0.45\textwidth]{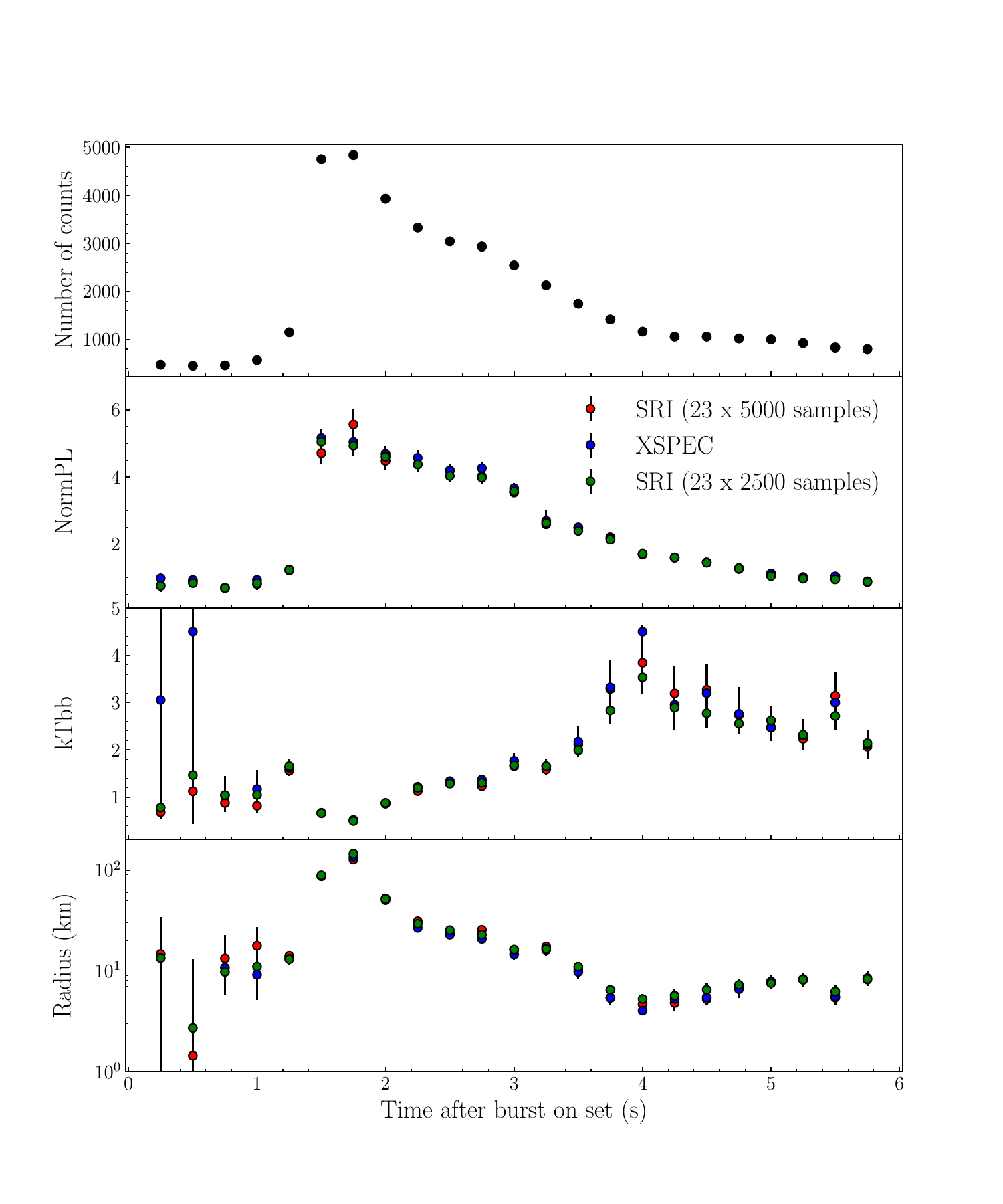}
    \caption{Left) 23 burst spectra covered by the restricted prior, indicated by the region in grey. Right) Recovered spectral parameters with SBI-NPE and XSPEC. The agreement between the different methods is remarkable. The figure shows the number of counts per spectrum (top), the normalization of the power law, the temperature of the backbody in keV, and at the bottom the blackbody normalization translated to a radius (in km) assuming a distance to the source of 8 kpc. Single Round Inference is performed with a training sample of $23\times5000$ spectra, derived from a restrictor, constraining the number of counts in the spectra to be between 100 and 10000 counts (red filled circles). Single Round Inference is also performed with a training sample of $23\times2500$ spectra, generated from a restrictor build from a quick inference (green filled circles). XSPEC best fit results are shown with blue filled circles.}
    \label{fig:db_sd_f15}
\end{figure*}

As we want to use the power of amortized inference, we use two methods to define our training sample. First, we apply the classifier with the condition to keep the model parameters associated with spectra with a number of counts ranging between 100–10000, hence covering plainly the range of counts of the observed spectra (400-5000). From the restricted prior, we consider arbitrarily 5000 simulated spectra per observed spectrum so that the network is trained with $23\times5000=115000$ samples. Second, we perform a coarse inference over the full prior range, and for each of the 23 spectra, we set the restricted prior as the posterior conditioned at the corresponding spectrum. The training sample is limited to 10000 spectra for the quick and coarse inference of the 23 spectra. For each of the 23 restricted prior, we then generate 2500 simulated spectra so that their ensemble will be used to train the network, i.e. with $23\times2500=57500$ samples. The predictive check of this prior is shown in Fig.~\ref{fig:db_sd_f15}, which indicates that such a build up prior covers all the observed spectra. The training then takes about $\sim 15$ minutes. The generation of the posterior samples takes $\sim 20$ seconds. We then fit the data with Bayesian inference with \xspec. We then derive the errors on the fitted parameters using MCMC. As can be seen, SBI-NPE with amortized inference for the two different restricted priors can follow the spectral evolution along the burst, with an accuracy comparable to \xspec, even when the number of counts in the spectra goes down to a few hundred, deep into the Poisson regime. The results of our fits are fully consistent with those reported by \citep{Keek2018ApJ...856L..37K,Strohmayer2019ApJ...878L..27S,Yu2023arXiv231216420Y}. This provides further demonstration that SBI-NPE is applicable to real data, and that the power of amortization can still be used for multiple spectra showing wide variability. 
\section{Discussion}
\label{Discussion}
We have demonstrated the first working principles of SBI-NPE for X-ray spectral fitting for both simulated and real data. We have shown that not only it can recover the same best fit parameters as traditional X-ray fitting techniques, but delivers also healthy posteriors, comparable to those derived from Bayesian inference with \xspec\ and BXA. The method works equally well in the Gaussian and Poisson regimes, with uncertainties reflecting the statistical quality of the data. The existence of a known likelihood helps to demonstrate that the method is well calibrated. We have still performed recommended checks, such as  Simulation-Based Calibrations (SBC) \citep{Talts2018arXiv180406788T}: a procedure for validating inferences from Bayesian algorithms capable of generating posterior samples. SBC provides a (qualitative) view and a quantitative measure to check, whether the uncertainties of the posterior are well-balanced, i.e., neither over-confident nor under-confident. SBI-NPE as implemented here passed the SBC checks, as expected from the comparison of the posterior distributions with \xspec\ and BXA. 

We have shown SBI-NPE to be less sensitive to local false minima than classical minimization techniques as implemented in \xspec, consistent with the findings of \cite{Parker2022MNRAS.514.4061P}. We have shown that, although raw spectra can train the network, SBI-NPE can be coupled with Principal Component Analysis for reducing the dimensions of the data to train the network, offering potential speed improvements for the inference \citep{Parker2022MNRAS.514.4061P}. For the simple models considered here, no broadening of the posterior distribution is observed. This kind of approach can be extended to various dimension reduction methods, since SBI-NPE is not bound to any likelihood computation. The latter also means that SBI-NPE could apply when formulating a likelihood is non optimal, e.g. in the case of the analysis of multi-dimensional Poisson data of extended X-ray sources \citep{Peterson2004ApJ...615..545P}.

Multiple-round inference combined with a restricted prior is perfectly suited when dealing with a few observations. The power of amortization can also be used, even when the observations show large spectral variability, as demonstrated above. The consideration of a restricted prior, either on the interval of counts covered by the observation data sets or from a coarse fitting of the ensemble of observations with a small neural network used upfront, enables to define an efficient training sample, to cover the targeted observations. Note however that the use of a restricted prior can always be compensated by a larger sample size on an extended prior, with the penalty on the training time. The use of a classifier to restrict the range of priors, being easy to implement and running fast, can be coupled with standard X-ray fitting tools to increase their speed and decrease the risk to get trapped into local false minima.

Within the demonstration of the working principles of the technique, we have found that the training time of the neural network is shorter and/or comparable to state of the art fitting tools. Speeding up the training may be possible using larger computer resources, e.g. moving from a laptop to a  cluster. On the other hand, once the network has been trained, generating the posterior distribution is instantaneous, and orders of magnitude faster than traditional fitting. This means that SBI-NPE holds great potential for integration in pipelines of data processing for massive data sets. The range of applications of the method has yet to be explored, but the ability to process a large sample of observations with the same network offers the opportunity to use it to track instrument dis-functioning, calibration errors, etc. This will be investigated for the X-IFU instrument on Athena \citep{Barret2023ExA....55..373B}. 

 We are aware that we have demonstrated the working principles of SBI-NPE, considering simple models, spectra with a relatively small number of bins. The applicability of the technique to more sophistical models, higher resolution spectra, such as those that will be provided by X-IFU, will have to be demonstrated, although the alternative tools, such as \xspec\ and BXA, may have issues on their own in terms of processing time. Through this demonstration, we have already identified some aspects of the technique to keep investigating. For instance, an amortized network is applicable to an ensemble of spectra that must have the same binning/grouping, the same exposure time… The latter could be relaxed if one is not interested in the normalization of the model components (flux), but just the variations of the other parameters (e.g. the index of a power law). Alternatively, the posterior distributions of the normalizations of each of the additive components of the models could be scaled afterwards to account for the different integration times, with respect to the integration time of the simulated spectra used for the training. On the other hand, there is no easy work around the case of spectra having different number of bins. This will require some further investigation. Obviously, there are many cases when meaningful information can still be derived considering similar grouping and integration time for the spectra, as we have shown above in the case of a type I X-ray burst. Once the network has been trained, the generation of posteriors being instantaneous, SBI-NPE makes it possible to track model parameter variations on timescales much shorter than what is possible today with existing tools.
 
The quality of the training depends on the size of the training sample. Caution is therefore recommended when lowering the sample size to increase the inference speed. It should be stated again, that when using amortized inference, the time to train the network will always be compensated afterwards, by the order of magnitude faster generation of the posteriors for a large number of spectra. There is no rule yet to define the minimum sample size. Similarly, for multiple-round inference, the number of iterations is a free parameter, that may need some fine-tuning to ensure that convergence to the best fit solution has been reached. The existence of alternative robust tools such as \xspec\ and BXA will help to define guidelines. For complex multi-component spectra, the size of the training sample will have to be increased. The use of a restricted prior will always help, but considering dimension reduction, e.g. decomposition in principal components, or the use of an embedding network to extract the relevant summary features of the data, may become mandatory. Any loss of information will have an impact on the inference itself. Another limitation may come from the time to generate the simulations to train the network. Here we have used simple models and a fully parallel version of the \xspec\ fakeit developed within \jaxspec\ (Dupourqué \& al. in preparation). This was not a limiting factor for this work, as generating thousands of simulations takes only a few seconds.  

The python scripts from which the results have been derived are based on the \texttt{sbi} package \citep{Tejero2020JOSS....5.2505T}. With this paper, we release through GitHub, the \sixsa\footnote{\url{https://github.com/dbxifu/SIXSA}} (Simulation-based Inference for X-ray Spectral Analysis) python package, from which the working principles of single and multiple round inference with neural posterior estimation have been demonstrated. The python scripts that come with reference spectra are straightforward to use, and can be customized for different applications. We also release a first version of \jaxspec\ to further support the development and use of SBI-NPE for X-ray spectral fitting. Note that \jaxspec\ is currently limited in the number of models available (only basic analytical models, as the ones used here, are available). It remains obviously possible to generate and use simulated spectra produced with software packages like \xspec. This is encouraged, in particular for testing SBI-NPE with a large number of model parameters.
\section{Conclusions and way forward}
\label{Conclusions}
We have demonstrated for the first time the working principles of fitting X-ray spectra with simulation-based inference with neural posterior estimation, down to the Poisson regime. We have applied the technique to real data, and demonstrated that SBI-NPE converges to the same best fit parameters and provides fitting errors comparable to Bayesian inference. We may therefore be at the eve of a new era for X-ray spectral fitting, but more work is needed to demonstrate the wider applicability of the technique to more sophisticated models and higher resolution spectra, and in particular to those provided by the new generation of instruments, such as the X-IFU spectrometer to fly on-board Athena. Yet, along this work, we have not identified any show-stoppers for this not to be achievable. Certainly, the pace at which machine learning applications develop across so many fields will also help in solving any issues that we may have to face, strengthening the case for developing further the potential of Simulation-Based Inference with Neural Posterior Estimation for X-ray spectral fitting. May the release of the \sixsa\ python package help the community to contribute to this exciting prospect. 
\section{Acknowledgments}
DB would like to thank all the colleagues who shared their unfortunate experience of getting stuck, without knowing, into local false minima, when doing X-ray spectral fitting. The fear of ignoring that the true global minimum is just nearby is what motivated this work in the first place with the hope of preventing sleepless nights in the future. DB is also grateful to all his X-IFU colleagues, in particular from CNES, for developing such a beautiful instrument that will require new tools, such as the one introduced here, for analyzing the high quality data that it will generate. DB/SD thank Alexei Molin and Erwan Quintin for their support and encouragements along this work. The authors are grateful to Fabio Acero, Maggie Lieu, the anonymous referee and the editor for useful comments on the paper. Finally DB thanks Michael Deistler for support in using and optimizing the restricted prior.

In addition to the \texttt{sbi} package \citep{Tejero2020JOSS....5.2505T}, this work made use of many awesome Python packages: \texttt{ChainConsumer} \citep{Hinton2016}, \texttt{matplotlib} \citep{matplotlib}, \texttt{numpy} \citep{numpy}, \texttt{pandas} \citep{pandas}, \texttt{pytorch} \citep{pytorch}, \texttt{scikit-learn} \citep{scikit-learn}, \texttt{scypi} \citep{scipy}, \texttt{tensorflow} \citep{tensorflow}.

The \sixsa\ Python package is available at this URL: \url{https://github.com/dbxifu/SIXSA}.

\bibliographystyle{aa} 
\bibliography{dbarret.bib} 

\begin{thebibliography}{47}
\expandafter\ifx\csname natexlab\endcsname\relax\def\natexlab#1{#1}\fi

\bibitem[{Abadi {et~al.}(2015)Abadi, Agarwal, Barham, Brevdo, Chen, Citro,
  Corrado, Davis, Dean, Devin, Ghemawat, Goodfellow, Harp, Irving, Isard, Jia,
  Jozefowicz, Kaiser, Kudlur, Levenberg, Man\'{e}, Monga, Moore, Murray, Olah,
  Schuster, Shlens, Steiner, Sutskever, Talwar, Tucker, Vanhoucke, Vasudevan,
  Vi\'{e}gas, Vinyals, Warden, Wattenberg, Wicke, Yu, \& Zheng}]{tensorflow}
Abadi, M., Agarwal, A., Barham, P., {et~al.} 2015, {TensorFlow}: Large-Scale
  Machine Learning on Heterogeneous Systems, software available from
  tensorflow.org

\bibitem[{{Arnaud}(1996)}]{Arnaud1996ASPC..101...17A}
{Arnaud}, K.~A. 1996, in Astronomical Society of the Pacific Conference Series,
  Vol. 101, Astronomical Data Analysis Software and Systems V, ed. G.~H.
  {Jacoby} \& J.~{Barnes}, 17

\bibitem[{{Barret} {et~al.}(2023){Barret}, {Albouys}, {Herder}, {Piro},
  {Cappi}, {Huovelin}, {Kelley}, {Mas-Hesse}, {Paltani}, {Rauw}, {Rozanska},
  {Svoboda}, {Wilms}, {Yamasaki}, {Audard}, {Bandler}, {Barbera}, {Barcons},
  {Bozzo}, {Ceballos}, {Charles}, {Costantini}, {Dauser}, {Decourchelle},
  {Duband}, {Duval}, {Fiore}, {Gatti}, {Goldwurm}, {Hartog}, {Jackson},
  {Jonker}, {Kilbourne}, {Korpela}, {Macculi}, {Mendez}, {Mitsuda}, {Molendi},
  {Pajot}, {Pointecouteau}, {Porter}, {Pratt}, {Pr{\^e}le}, {Ravera}, {Sato},
  {Schaye}, {Shinozaki}, {Skup}, {Soucek}, {Thibert}, {Vink}, {Webb}, {Chaoul},
  {Raulin}, {Simionescu}, {Torrejon}, {Acero}, {Branduardi-Raymont}, {Ettori},
  {Finoguenov}, {Grosso}, {Kaastra}, {Mazzotta}, {Miller}, {Miniutti},
  {Nicastro}, {Sciortino}, {Yamaguchi}, {Beaumont}, {Cucchetti}, {D'Andrea},
  {Eckart}, {Ferrando}, {Kammoun}, {Lotti}, {Mesnager}, {Natalucci}, {Peille},
  {de Plaa}, {Ardellier}, {Argan}, {Bellouard}, {Carron}, {Cavazzuti},
  {Fiorini}, {Khosropanah}, {Martin}, {Perry}, {Pinsard}, {Pradines}, {Rigano},
  {Roelfsema}, {Schwander}, {Torrioli}, {Ullom}, {Vera}, {Villegas},
  {Zuchniak}, {Brachet}, {Cicero}, {Doriese}, {Durkin}, {Fioretti}, {Geoffray},
  {Jacques}, {Kirsch}, {Smith}, {Adams}, {Gloaguen}, {Hoogeveen}, {van der
  Hulst}, {Kiviranta}, {van der Kuur}, {Ledot}, {van Leeuwen}, {van Loon},
  {Lyautey}, {Parot}, {Sakai}, {van Weers}, {Abdoelkariem}, {Adam}, {Adami},
  {Aicardi}, {Akamatsu}, {Alonso}, {Amato}, {Andr{\'e}}, {Angelinelli},
  {Anon-Cancela}, {Anvar}, {Atienza}, {Attard}, {Auricchio}, {Balado},
  {Bancel}, {Barusso}, {Bascu{\~n}an}, {Bernard}, {Berrocal}, {Blin}, {Bonino},
  {Bonnet}, {Bonny}, {Boorman}, {Boreux}, {Bounab}, {Boutelier}, {Boyce},
  {Brienza}, {Bruijn}, {Bulgarelli}, {Calarco}, {Callanan}, {Campello},
  {Camus}, {Canourgues}, {Capobianco}, {Cardiel}, {Castellani}, {Cheatom},
  {Chervenak}, {Chiarello}, {Clerc}, {Clerc}, {Cobo}, {Coeur-Joly}, {Coleiro},
  {Colonges}, {Corcione}, {Coriat}, {Coynel}, {Cuttaia}, {D'Ai}, {D'anca},
  {Dadina}, {Daniel}, {Dauner}, {DeNigris}, {Dercksen}, {DiPirro}, {Doumayrou},
  {Dubbeldam}, {Dupieux}, {Dupourqu{\'e}}, {Durand}, {Eckert}, {Eiriz},
  {Ercolani}, {Etcheverry}, {Finkbeiner}, {Fiocchi}, {Fossecave}, {Franssen},
  {Frericks}, {Gabici}, {Gant}, {Gao}, {Gastaldello}, {Genolet}, {Ghizzardi},
  {Gil}, {Giovannini}, {Godet}, {Gomez-Elvira}, {Gonzalez}, {Gonzalez},
  {Gottardi}, {Granat}, {Gros}, {Guignard}, {Hieltjes}, {Hurtado}, {Irwin},
  {Jacquey}, {Janiuk}, {Jaubert}, {Jim{\'e}nez}, {Jolly}, {Jourdan}, {Julien},
  {Kedziora}, {Korb}, {Kreykenbohm}, {K{\"o}nig}, {Langer}, {Laudet},
  {Laurent}, {Laurenza}, {Lesrel}, {Ligori}, {Lorenz}, {Luminari}, {Maffei},
  {Maisonnave}, {Marelli}, {Massonet}, {Maussang}, {Melchor}, {Le Mer},
  {Millan}, {Millerioux}, {Mineo}, {Minervini}, {Molin}, {Monestes},
  {Montinaro}, {Mot}, {Murat}, {Nagayoshi}, {Naz{\'e}}, {Nogu{\`e}s}, {Pailot},
  {Panessa}, {Parodi}, {Petit}, {Piconcelli}, {Pinto}, {Plaza}, {Plaza},
  {Poyatos}, {Prouv{\'e}}, {Ptak}, {Puccetti}, {Puccio}, {Ramon}, {Reina},
  {Rioland}, {Rodriguez}, {Roig}, {Rollet}, {Roncarelli}, {Roudil}, {Rudnicki},
  {Sanisidro}, {Sciortino}, {Silva}, {Sordet}, {Soto-Aguilar}, {Spizzi},
  {Surace}, {Fern{\'a}ndez S{\'a}nchez}, {Taralli}, {Terrasa}, {Terrier},
  {Todaro}, {Ubertini}, {Uslenghi}, {de Vaate}, {Vaccaro}, {Varisco},
  {Varni{\`e}re}, {Vibert}, {Vidriales}, {Villa}, {Vodopivec}, {Volpe}, {de
  Vries}, {Wakeham}, {Walmsley}, {Wise}, {de Wit}, \&
  {Wo{\'z}niak}}]{Barret2023ExA....55..373B}
{Barret}, D., {Albouys}, V., {Herder}, J.-W.~d., {et~al.} 2023, Experimental
  Astronomy, 55, 373

\bibitem[{Bevington \& Robinson(2003)}]{bevington2003data}
Bevington, P. \& Robinson, D. 2003, Data Reduction and Error Analysis for the
  Physical Sciences (McGraw-Hill Education)

\bibitem[{Bradbury {et~al.}(2018)Bradbury, Frostig, Hawkins, Johnson, Leary,
  Maclaurin, Necula, Paszke, Vander{P}las, Wanderman-{M}ilne, \&
  Zhang}]{jax2018github}
Bradbury, J., Frostig, R., Hawkins, P., {et~al.} 2018, {JAX}: composable
  transformations of {P}ython+{N}um{P}y programs

\bibitem[{Buchner(2021)}]{Buchner2021Ultranest}
Buchner, J. 2021, Journal of Open Source Software, 6, 3001

\bibitem[{{Buchner} \& {Boorman}(2023)}]{Buchner2023arXiv230905705B}
{Buchner}, J. \& {Boorman}, P. 2023, arXiv e-prints, arXiv:2309.05705

\bibitem[{{Buchner} {et~al.}(2014){Buchner}, {Georgakakis}, {Nandra}, {Hsu},
  {Rangel}, {Brightman}, {Merloni}, {Salvato}, {Donley}, \&
  {Kocevski}}]{Buchner2014A&A...564A.125B}
{Buchner}, J., {Georgakakis}, A., {Nandra}, K., {et~al.} 2014, \aap, 564, A125

\bibitem[{{Cash}(1979)}]{Cash1979ApJ...228..939C}
{Cash}, W. 1979, \apj, 228, 939

\bibitem[{{Cranmer} {et~al.}(2020){Cranmer}, {Brehmer}, \&
  {Louppe}}]{Cranmer2020PNAS..11730055C}
{Cranmer}, K., {Brehmer}, J., \& {Louppe}, G. 2020, Proceedings of the National
  Academy of Science, 117, 30055

\bibitem[{{Crisostomi} {et~al.}(2023){Crisostomi}, {Dey}, {Barausse}, \&
  {Trotta}}]{Crisostomi2023PhRvD.108d4029C}
{Crisostomi}, M., {Dey}, K., {Barausse}, E., \& {Trotta}, R. 2023, \prd, 108,
  044029

\bibitem[{{Deistler} {et~al.}(2022){Deistler}, {Goncalves}, \&
  {Macke}}]{Deistler2022arXiv221004815D}
{Deistler}, M., {Goncalves}, P.~J., \& {Macke}, J.~H. 2022, arXiv e-prints,
  arXiv:2210.04815

\bibitem[{Deistler {et~al.}(2022)Deistler, Macke, \&
  Gonçalves}]{deistler2022pnas2207632119}
Deistler, M., Macke, J.~H., \& Gonçalves, P.~J. 2022, Proceedings of the
  National Academy of Sciences, 119, e2207632119

\bibitem[{{Gendreau} {et~al.}(2012){Gendreau}, {Arzoumanian}, \&
  {Okajima}}]{Gendreau2012SPIE.8443E..13G}
{Gendreau}, K.~C., {Arzoumanian}, Z., \& {Okajima}, T. 2012, in Society of
  Photo-Optical Instrumentation Engineers (SPIE) Conference Series, Vol. 8443,
  Space Telescopes and Instrumentation 2012: Ultraviolet to Gamma Ray, ed.
  T.~{Takahashi}, S.~S. {Murray}, \& J.-W.~A. {den Herder}, 844313

\bibitem[{Germain {et~al.}(2015)Germain, Gregor, Murray, \&
  Larochelle}]{germainMADEMaskedAutoencoder2015}
Germain, M., Gregor, K., Murray, I., \& Larochelle, H. 2015, in Proceedings of
  the 32nd {International} {Conference} on {Machine} {Learning} (PMLR),
  881--889, iSSN: 1938-7228

\bibitem[{{Goodman} \& {Weare}(2010)}]{2010CAMCS...5...65G}
{Goodman}, J. \& {Weare}, J. 2010, Communications in Applied Mathematics and
  Computational Science, 5, 65

\bibitem[{{Graber} {et~al.}(2023){Graber}, {Ronchi}, {Pardo-Araujo}, \&
  {Rea}}]{Graber2023arXiv231214848G}
{Graber}, V., {Ronchi}, M., {Pardo-Araujo}, C., \& {Rea}, N. 2023, arXiv
  e-prints, arXiv:2312.14848

\bibitem[{{Greenberg} {et~al.}(2019){Greenberg}, {Nonnenmacher}, \&
  {Macke}}]{Greenberg2019arXiv190507488G}
{Greenberg}, D.~S., {Nonnenmacher}, M., \& {Macke}, J.~H. 2019, arXiv e-prints,
  arXiv:1905.07488

\bibitem[{Greenberg {et~al.}(2019)Greenberg, Nonnenmacher, \&
  Macke}]{greenbergAutomaticPosteriorTransformation2019}
Greenberg, D.~S., Nonnenmacher, M., \& Macke, J.~H. 2019, Automatic {Posterior}
  {Transformation} for {Likelihood}-{Free} {Inference}, arXiv:1905.07488 [cs,
  stat]

\bibitem[{Harris {et~al.}(2020)Harris, Millman, van~der Walt, Gommers,
  Virtanen, Cournapeau, Wieser, Taylor, Berg, Smith, Kern, Picus, Hoyer, van
  Kerkwijk, Brett, Haldane, del R{\'{i}}o, Wiebe, Peterson,
  G{\'{e}}rard-Marchant, Sheppard, Reddy, Weckesser, Abbasi, Gohlke, \&
  Oliphant}]{numpy}
Harris, C.~R., Millman, K.~J., van~der Walt, S.~J., {et~al.} 2020, Nature, 585,
  357

\bibitem[{He {et~al.}(2015)He, Zhang, Ren, \& Sun}]{he2015deep}
He, K., Zhang, X., Ren, S., \& Sun, J. 2015, Deep Residual Learning for Image
  Recognition

\bibitem[{{Hinton}(2016)}]{Hinton2016}
{Hinton}, S.~R. 2016, The Journal of Open Source Software, 1, 00045

\bibitem[{Hoffman \& Gelman(2011)}]{hoffman2011nouturn}
Hoffman, M.~D. \& Gelman, A. 2011, The No-U-Turn Sampler: Adaptively Setting
  Path Lengths in Hamiltonian Monte Carlo

\bibitem[{Hunter(2007)}]{matplotlib}
Hunter, J.~D. 2007, Computing in science \& engineering, 9, 90

\bibitem[{{Huppenkothen} \&
  {Bachetti}(2022)}]{Huppenkothen_2022MNRAS.511.5689H}
{Huppenkothen}, D. \& {Bachetti}, M. 2022, \mnras, 511, 5689

\bibitem[{{Ichinohe} {et~al.}(2018){Ichinohe}, {Yamada}, {Miyazaki}, \&
  {Saito}}]{Ichinohe2018MNRAS.475.4739I}
{Ichinohe}, Y., {Yamada}, S., {Miyazaki}, N., \& {Saito}, S. 2018, \mnras, 475,
  4739

\bibitem[{{Kaastra}(2017)}]{Kaastra2017A&A...605A..51K}
{Kaastra}, J.~S. 2017, \aap, 605, A51

\bibitem[{{Keek} {et~al.}(2018){Keek}, {Arzoumanian}, {Chakrabarty},
  {Chenevez}, {Gendreau}, {Guillot}, {G{\"u}ver}, {Homan}, {Jaisawal},
  {LaMarr}, {Lamb}, {Mahmoodifar}, {Markwardt}, {Okajima}, {Strohmayer}, \& {in
  't Zand}}]{Keek2018ApJ...856L..37K}
{Keek}, L., {Arzoumanian}, Z., {Chakrabarty}, D., {et~al.} 2018, \apjl, 856,
  L37

\bibitem[{{Khullar} {et~al.}(2022){Khullar}, {Nord}, {{\'C}iprijanovi{\'c}},
  {Poh}, \& {Xu}}]{Khullar2022MLS&T...3dLT04K}
{Khullar}, G., {Nord}, B., {{\'C}iprijanovi{\'c}}, A., {Poh}, J., \& {Xu}, F.
  2022, Machine Learning: Science and Technology, 3, 04LT04

\bibitem[{Kobyzev {et~al.}(2021)Kobyzev, Prince, \&
  Brubaker}]{kobyzevNormalizingFlowsIntroduction2021}
Kobyzev, I., Prince, S. J.~D., \& Brubaker, M.~A. 2021, IEEE Transactions on
  Pattern Analysis and Machine Intelligence, 43, 3964, arXiv:1908.09257 [cs,
  stat]

\bibitem[{{Lueckmann} {et~al.}(2017){Lueckmann}, {Goncalves}, {Bassetto},
  {{\"O}cal}, {Nonnenmacher}, \& {Macke}}]{Lueckmann2017arXiv171101861L}
{Lueckmann}, J.-M., {Goncalves}, P.~J., {Bassetto}, G., {et~al.} 2017, arXiv
  e-prints, arXiv:1711.01861

\bibitem[{McKinney(2010)}]{pandas}
McKinney, W. 2010, in Proceedings of the 9th Python in Science Conference, ed.
  S.~van~der Walt \& J.~Millman, 51 -- 56

\bibitem[{{Papamakarios} \& {Murray}(2016)}]{Papamakarios2016arXiv160506376P}
{Papamakarios}, G. \& {Murray}, I. 2016, arXiv e-prints, arXiv:1605.06376

\bibitem[{Papamakarios {et~al.}(2021)Papamakarios, Nalisnick, Rezende, Mohamed,
  \& Lakshminarayanan}]{papamakariosNormalizingFlowsProbabilistic2021}
Papamakarios, G., Nalisnick, E., Rezende, D.~J., Mohamed, S., \&
  Lakshminarayanan, B. 2021, Normalizing {Flows} for {Probabilistic} {Modeling}
  and {Inference}, arXiv:1912.02762 [cs, stat]

\bibitem[{Papamakarios {et~al.}(2017)Papamakarios, Pavlakou, \&
  Murray}]{papamakariosMaskedAutoregressiveFlow2017}
Papamakarios, G., Pavlakou, T., \& Murray, I. 2017, in Advances in {Neural}
  {Information} {Processing} {Systems}, Vol.~30 (Curran Associates, Inc.)

\bibitem[{{Parker} {et~al.}(2022){Parker}, {Lieu}, \&
  {Matzeu}}]{Parker2022MNRAS.514.4061P}
{Parker}, M.~L., {Lieu}, M., \& {Matzeu}, G.~A. 2022, \mnras, 514, 4061

\bibitem[{Paszke {et~al.}(2017)Paszke, Gross, Chintala, Chanan, Yang, DeVito,
  Lin, Desmaison, Antiga, \& Lerer}]{pytorch}
Paszke, A., Gross, S., Chintala, S., {et~al.} 2017, in NIPS Autodiff Workshop

\bibitem[{Pedregosa {et~al.}(2011)Pedregosa, Varoquaux, Gramfort, Michel,
  Thirion, Grisel, Blondel, Prettenhofer, Weiss, Dubourg, Vanderplas, Passos,
  Cournapeau, Brucher, Perrot, \& Duchesnay}]{scikit-learn}
Pedregosa, F., Varoquaux, G., Gramfort, A., {et~al.} 2011, Journal of Machine
  Learning Research, 12, 2825

\bibitem[{{Peterson} {et~al.}(2004){Peterson}, {Jernigan}, \&
  {Kahn}}]{Peterson2004ApJ...615..545P}
{Peterson}, J.~R., {Jernigan}, J.~G., \& {Kahn}, S.~M. 2004, \apj, 615, 545

\bibitem[{{Strohmayer} {et~al.}(2019){Strohmayer}, {Altamirano}, {Arzoumanian},
  {Bult}, {Chakrabarty}, {Chenevez}, {Fabian}, {Gendreau}, {Guillot}, {in 't
  Zand}, {Jaisawal}, {Keek}, {Kosec}, {Ludlam}, {Mahmoodifar}, {Malacaria}, \&
  {Miller}}]{Strohmayer2019ApJ...878L..27S}
{Strohmayer}, T.~E., {Altamirano}, D., {Arzoumanian}, Z., {et~al.} 2019, \apjl,
  878, L27

\bibitem[{{Talts} {et~al.}(2018){Talts}, {Betancourt}, {Simpson}, {Vehtari}, \&
  {Gelman}}]{Talts2018arXiv180406788T}
{Talts}, S., {Betancourt}, M., {Simpson}, D., {Vehtari}, A., \& {Gelman}, A.
  2018, arXiv e-prints, arXiv:1804.06788

\bibitem[{{Tejero-Cantero} {et~al.}(2020){Tejero-Cantero}, {Boelts},
  {Deistler}, {Lueckmann}, {Durkan}, {Gon{\c{c}}alves}, {Greenberg}, \&
  {Macke}}]{Tejero2020JOSS....5.2505T}
{Tejero-Cantero}, A., {Boelts}, J., {Deistler}, M., {et~al.} 2020, The Journal
  of Open Source Software, 5, 2505

\bibitem[{{Vasist} {et~al.}(2023){Vasist}, {Rozet}, {Absil}, {Molli{\`e}re},
  {Nasedkin}, \& {Louppe}}]{Vasist2023A&A...672A.147V}
{Vasist}, M., {Rozet}, F., {Absil}, O., {et~al.} 2023, \aap, 672, A147

\bibitem[{{Verner} {et~al.}(1996){Verner}, {Ferland}, {Korista}, \&
  {Yakovlev}}]{Verner_1996}
{Verner}, D.~A., {Ferland}, G.~J., {Korista}, K.~T., \& {Yakovlev}, D.~G. 1996,
  \apj, 465, 487

\bibitem[{{Virtanen} {et~al.}(2020){Virtanen}, {Gommers}, {Oliphant},
  {Haberland}, {Reddy}, {Cournapeau}, {Burovski}, {Peterson}, {Weckesser},
  {Bright}, {van der Walt}, {Brett}, {Wilson}, {Jarrod Millman}, {Mayorov},
  {Nelson}, {Jones}, {Kern}, {Larson}, {Carey}, {Polat}, {Feng}, {Moore},
  {VanderPlas}, {Laxalde}, {Perktold}, {Cimrman}, {Henriksen}, {Quintero},
  {Harris}, {Archibald}, {Ribeiro}, {Pedregosa}, {van Mulbregt}, \&
  {Contributors}}]{scipy}
{Virtanen}, P., {Gommers}, R., {Oliphant}, T.~E., {et~al.} 2020, Nature Methods

\bibitem[{{Wilms} {et~al.}(2000){Wilms}, {Allen}, \&
  {McCray}}]{Wilms2000ApJ...542..914W}
{Wilms}, J., {Allen}, A., \& {McCray}, R. 2000, \apj, 542, 914

\bibitem[{{Yu} {et~al.}(2023){Yu}, {Li}, {Lu}, {Pan}, {Yang}, {Chen}, {Zhang},
  \& {Falanga}}]{Yu2023arXiv231216420Y}
{Yu}, W., {Li}, Z., {Lu}, Y., {et~al.} 2023, arXiv e-prints, arXiv:2312.16420

\end{thebibliography}

\end{document}